\documentclass[aps,twocolumn,superscriptaddress,floatfix,noeprint, pra]{revtex4-2}
\usepackage{algorithm}
\usepackage{algpseudocode}
\usepackage{blindtext}
\usepackage{bm}
\usepackage{braket}
\usepackage{amssymb}
\usepackage{amsmath}
\usepackage{graphicx}
\usepackage{float}
\usepackage{bbold}
\usepackage{xcolor}
\usepackage{mathtools}
\usepackage{mathrsfs}
\usepackage[normalem]{ulem}

\usepackage[breaklinks=true,colorlinks=true,linkcolor=teal,urlcolor=teal,citecolor=teal]{hyperref}
\usepackage{autonum}

\algrenewcommand\algorithmicrequire{\textbf{Input:}}
\algrenewcommand\algorithmicensure{\textbf{Output:}}

\renewcommand{\t}[1]{\textrm{#1}}
\newcommand{\Lmb}{\overset{\leftrightarrow}{\Lambda}}
\newcommand{\varLmb}{\overset{\leftrightarrow}{\varLambda}}

\begin{document}
\title{Universal bounds for quantum metrology in the presence of correlated noise}
\author{Stanis{\l}aw Kurdzia{\l}ek}
\affiliation{Faculty of Physics, University of Warsaw, Pasteura 5, 02-093 Warszawa, Poland}
\author{Francesco Albarelli}
\affiliation{Scuola Normale Superiore, I-56126 Pisa, Italy}
\author{Rafa{\l}  Demkowicz-Dobrza{\'n}ski} 
\affiliation{Faculty of Physics, University of Warsaw, Pasteura 5, 02-093 Warszawa, Poland}

\begin{abstract}
We derive fundamental bounds for general quantum metrological models involving both temporal or spatial correlations (mathematically described by quantum combs), which may be effectively computed in the limit of a large number of probes or sensing channels involved.
Although the bounds are not guaranteed to be tight in general, their tightness may be systematically increased by increasing numerical complexity of the procedure.
Interestingly, this approach yields bounds tighter than the state of the art also for uncorrelated channels.
We apply the bound to study the limits for the most general adaptive phase estimation models in the presence of temporally correlated dephasing.
We consider dephasing both parallel (no Heisenberg scaling)  and perpendicular (Heisenberg scaling possible) to the signal.
In the former case our new bounds show that negative correlations are beneficial, for the latter we show evidence that the bounds are tight.
We also apply the bounds to collisional thermometry, i.e. estimation of a parameter of the environment, showing evidence that entangled probes may provide only a limited advantage.
\end{abstract}

\maketitle

\paragraph*{Introduction.}
Reducing noise effects is the foremost challenge in advancing quantum technologies~\cite{Suter2015}.
Real-world quantum systems suffer from environmental interactions, leading to decoherence and particle loss, yet quantum error correction (QEC) codes often help to achieve quantum advantage~\cite{Lidar2013}.

In quantum metrology~\cite{Giovannetti2004,Giovannetti2011,Demkowicz-Dobrzanski2015a,Degen2016,Pezze2018,Pirandola2018,Polino2020,Liu2022c,Jiao2023a,Liu2024s,Montenegro2024}, the most spectacular quantum advantage is the Heisenberg scaling (HS), a precision that scales as $1/N$, where $N$ is the number of resources (such as particles or channel uses), whereas for standard scaling (SS) the precision is proportional to $1/\sqrt{N}$.
Strategies that attain the HS are known for noiseless~\cite{Ou1997, Giovannetti2004} and noisy~\cite{Kessler2014, Dur2014, Zhou2017} phase estimation---in the latter case, proper QEC is indispensable.

Recent developments in quantum channel estimation theory enable full characterization of precision limits under \emph{uncorrelated} noise, allowing one to determine the optimal  scaling (HS or SS) and the corresponding constant via a simple semidefinite program (SDP)~\cite{Fujiwara2008,Escher2011,Demkowicz-Dobrzanski2012,Koodynski2013,Demkowicz-Dobrzanski2014,Sekatski2016,Zhou2017,Demkowicz-Dobrzanski2017,Zhou2019e,Zhou2020,Kurdzialek2023a}.
However, noise and signal correlations are significant in many  systems~\cite{Banaszek2004,Caruso2014,deVega2017,Szankowski2017b,Pollock2018a,VonLupke2020,Milz2021,Harper2023,Mele2023,Preskill2013}.
Despite various case studies including temporal~\cite{Matsuzaki2011,Chin2012,Macieszczak2015,Smirne2015a,Haase2017,Szankowski2014,Smirne2018,Tamascelli2020,Altherr2021,Yang2024j} and spatial~\cite{Jeske2014,Layden2018,Czajkowski2019,Layden2019,Layden2020,Planella2022} correlations, or both~\cite{Beaudoin2018,Riberi2022},  metrological bounds for these cases are missing, especially in large-$N$ limit.
As such, one cannot assess the fundamental optimality of a specific protocol.
In this work, we fill this gap and derive universal bounds for correlated noise models.

\paragraph*{State-of-the-art bounds.}
Numerous metrological tasks, e.g. phase estimation or superresolution imaging~\cite{Tsang2016}, amount to the estimation of a single parameter $\theta$---the goal is to find a (locally) unbiased estimator $\tilde \theta$ with minimal variance $\Delta^2 \tilde \theta $.
When the parameter is encoded in a quantum state $\rho_\theta$, the \emph{quantum Cram\'er-Rao Bound} (QCRB) says that $\Delta^2 \tilde \theta \ge \frac{1}{F(\rho_\theta)}$, where $F(\rho_\theta)$ is the \emph{quantum Fisher information} (QFI)~\cite{helstrom1976quantum,Holevo2011b} (see 
 Appendix~\ref{app:QFI} for the definition).
The QFI is a local quantity that depends only on $\rho_\theta$ and its derivative $\dot \rho_\theta = \partial_\theta \rho_\theta$ evaluated at some specific value of $\theta$. 

Let $\ket{\Psi_\theta}$ be a purification of $\rho_\theta$, then it holds that $F(\ket{\Psi_\theta}) \ge F(\rho_\theta)$ because any measurement on $\rho_\theta$ can be implemented on $\ket{\Psi_\theta}$.
Moreover, the QFI is equal to a minimization over all purifications of $\rho_\theta$:
\begin{equation}
    \label{eq:QFI_pure}
        F(\rho_\theta)  = 4 \min_{\ket{\Psi_\theta} } \braket{\dot \Psi_\theta|\dot \Psi_\theta}, 
\end{equation}
which follows from the existence of a \textit{QFI non-increasing purification} (QFI NIP) satisfying~\cite{Fujiwara2008,Kolodynski2014} $F(\ket{\Psi_\theta}) = F(\rho_\theta)$.

In quantum metrology, the parameter $\theta$ is encoded in a quantum channel $\Lambda_\theta(\bullet) = \sum_k K_{k,\theta} \bullet K_{k,\theta}^\dagger$, where $K_{k,\theta}$ are Kraus operators \cite{Demkowicz-Dobrzanski2012}.
The ultimate precision of $\theta$ estimation is quantified by the \textit{channel} QFI $\mathcal{F}(\Lambda_\theta) = \max_{\rho} F(\rho_\theta)$, where $\rho_\theta = (\Lambda_\theta \otimes \mathcal{I}_\mathcal{A}) (\rho)$ and $\rho \in \mathcal{L}(\mathcal{H} \otimes \mathcal{A})$ is a possibly entangled input state of probe ($\mathcal{H}$) and noiseless ancillary ($\mathcal{A}$) systems; $\mathcal{L}(\bullet)$ is the set of linear operators acting on $\bullet$, and $\mathcal{I}$ the identity channel.
Starting from \eqref{eq:QFI_pure} one finds~\cite{Fujiwara2008}:
\begin{equation}
\label{eq:chan_QFI}
 \mathcal{F}(\Lambda_\theta) = 4 \min_{h} \| \alpha \|, \quad \alpha = \sum_k \dot {\tilde K}^\dagger_{k,\theta} \dot {\tilde K}_{k,\theta}, 
\end{equation}
where $\| \bullet \|$ is the operator norm, $\dot {\tilde K}_{k,\theta} = \dot { K}_{k,\theta} - i h_{kk'} K_{k',\theta}$ and $h$ is a hermitian matrix; the summation is performed over repeated indices. Importantly,  \eqref{eq:chan_QFI} can be recast as a simple SDP~\cite{Demkowicz-Dobrzanski2012}. 

The real potential of quantum metrology appears when $N>1$ channel copies are probed collectively using  \textit{adaptive} or \textit{active quantum feedback} (AD) strategies, where channels $\Lambda_\theta : \mathcal{L}(\mathcal{H}_{2i-1}) \rightarrow \mathcal{L}(\mathcal{H}_{2i})$ for $i \in \{1,2,...,N\}$ are probed sequentially, and arbitrary quantum control channels
$\t{C}_i: \mathcal{L} (\mathcal{H}_{2i} \otimes \mathcal{A}_i) \rightarrow \mathcal{L} (\mathcal{H}_{2i+1} \otimes \mathcal{A}_{i+1})$
can act on a probe and arbitrarily large ancilla $\mathcal{A}_i$ after each $\Lambda_\theta$.
The AD class covers all strategies besides those involving indefinite causal order~\cite{Liu2023b}---in particular \textit{parallel} strategies, i.e. all channels simultaneously probed by an entangled state, form a subset of AD~\cite{Demkowicz-Dobrzanski2014}.

A general AD strategy can be represented by a \textit{quantum comb}, which models the sequence of channels sharing a common environment \cite{Chiribella2008b}. Formally,  $ \t{E}\in \t{Comb}[(\mathcal{K}_1,\mathcal{K}_2),...,(\mathcal{K}_{2N-1}, \mathcal{K}_{2N})]$  when $\t{E}$ is a quantum channel with inputs $\mathcal{K}_1$, $\mathcal{K}_3$,... $\mathcal{K}_{2N-1}$ and outputs $\mathcal{K}_2$, $\mathcal{K}_4$,... $\mathcal{K}_{2N}$, satisfying some extra linear conditions (see Appendix~\ref{appB}) ensuring that output $\mathcal{K}_{2j}$ may only depend on input $\mathcal{K}_{2k-1}$ when $j \le k$. 
The Hilbert spaces $\mathcal{K}_{2k-1}, \mathcal{K}_{2k}$ can be interpreted as input/output of $k$th comb tooth respectively. 

\begin{figure*}[t]
\includegraphics[width = 0.9 \linewidth]{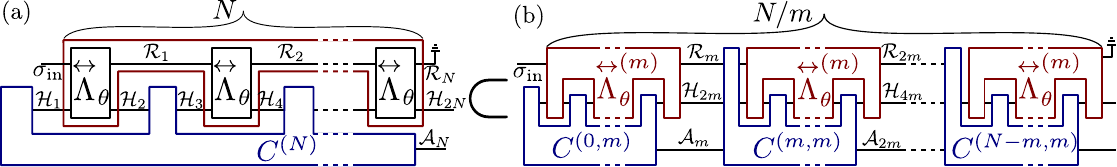}
\caption{The comb $\Lambda_\theta^{(N)}$ consists of teeth $\Lmb_\theta$ connected with environmental spaces $\mathcal{R}$; physical estimation strategies $C^{(N)}$ act only on probe subspaces $\mathcal{H}$ (a).
To derive an upper bound, we divide  $\Lambda_\theta^{(N)}$ into blocks of $m$ teeth $\Lmb_\theta^{(m)}$ (b), and allow for more powerful strategies, in which a purified state of $\mathcal{R}, \mathcal{H}, \mathcal{A}$ after each block is an input of the next control $\t{C}^{(l,m)}$.
}
\label{fig:corr} 
\end{figure*}

Input state $\rho \in \mathcal{L}(\mathcal{H}_1)$ and controls $\t{C}_i$  connected with ancillae $\mathcal{A}_i$ are  represented using a comb $\t{C}^{(N)} \in \text{Comb} [(\emptyset , \mathcal{H}_1) , (\mathcal{H}_2, \mathcal{H}_3), ... , ( \mathcal{H}_{2N-2}, \mathcal{H}_{2N-1} \otimes \mathcal{A}_{N})]$, where $\emptyset$ is a trivial space (no input), see Fig.~\ref{fig:corr}(a). 
The parameter encoding is described by a comb $\Lambda_\theta^{(N)} \in \text{Comb} [( \mathcal{H}_1, \mathcal{H}_2) ,  ... , ( \mathcal{H}_{2N-1}, \mathcal{H}_{2N}  )]$ which can model any type of noise and signal correlations; for uncorrelated channels it reduces to $\Lambda_\theta^{(N)} = \Lambda_\theta^{\otimes N}$.
The output state $\rho_\theta \in \mathcal{L}(\mathcal{H}_{2N} \otimes \mathcal{A}_N)$ is obtained by concatenating the corresponding inputs and outputs of $\t{C}^{(N)}$ and $\Lambda_\theta^{(N)}$ through the \textit{link product} operation~\cite{Chiribella2008b, Chiribella2009}, $\rho_\theta =\t{C}^{(N)} \star \Lambda_\theta^{(N)}$.

The  \textit{comb QFI}~\cite{Altherr2021} $\mathcal{F}_\text{AD}^{(N)} = \max_{\t{C}^{(N)} } F(\t{C}^{(N)} \star \Lambda_\theta^{(N)})$ quantifies the ultimate estimation precision for a parameter encoded in a comb $\Lambda_\theta^{(N)}$. To evaluate $\mathcal{F}_\text{AD}^{(N)}$, we should decompose the Choi-Jamiołkowski (CJ)  operator \footnote{The CJ operator of a channel $\Lambda: \mathcal{H}_\text{in} \rightarrow \mathcal{H}_\text{out}$ is defined as $\varLambda = \Lambda \otimes \mathcal{I} (\ket{\Psi_+} \bra{\Psi_+})$, where $\ket{\Psi_+} = \sum_i \ket{i} \ket{i}$ , $\ket{i}$ is orthonormal basis of $\mathcal{H}_\text{in}$, $\mathcal{I}$ is and identity channel. } of $\Lambda_\theta^{(N)}$
as $\varLambda_\theta^{(N)} = \sum_k \ket{K_{k,\theta}^{(N)}}\bra{K_{k,\theta}^{(N)}}$, where 
$\ket{K_{k,\theta}^{(N)}}$ are vectorized Kraus operators; we use roman font for channels, and italics for the corresponding CJ operators. 
Analogously to $\alpha$ in \eqref{eq:chan_QFI}, a \textit{performance operator} is defined as $\boldsymbol{\alpha}^{(N)} = \text{Tr}_\text{out} \left( \sum_k \ket{\dot {\tilde K}_{k,\theta}^{(N)}} \bra{\dot{\tilde K}_{k,\theta}^{(N)}} \right)$ \footnote{The performance operator is often defined as transposition of this expression \cite{Altherr2021}. It does not affect the further results---for example, \eqref{eq:comb_QFI} remains valid irrespectively of the convention because $(C^{(N)})^T$ is a comb iff $C^{(N)}$ is a comb. The typical notation for performance operator is $\Omega$, here we use $\boldsymbol{\alpha}$ because of analogy with 
 a single-channel object. } , where $\text{Tr}_\text{out}$ is the partial trace over the last output space of $\varLambda_\theta^{(N)}$ ($\mathcal{H}_{2N}$),  $\ket{\dot {\tilde K}_{k,\theta}^{(N)}} = \ket{\dot { K}_{k,\theta}^{(N)}} - i h_{kk'} \ket{{ K}_{k',\theta}^{(N)}} $, $h$ is a hermitian matrix.
Thus, the comb QFI
can be rewritten as \cite{Altherr2021, Liu2023b}
\begin{equation}
\label{eq:comb_QFI}
    \mathcal{F}_\text{AD}^{(N)} = 4 \min_h \max_{\tilde C^{(N)}} \text{Tr} \left(  \boldsymbol{\alpha}^{(N)} \tilde C^{(N)} \right),
\end{equation}
where $\tilde C^{(N)} = \text{Tr}_{\mathcal{A}_N} C^{(N)}$.
This optimization can be formulated as an SDP ~\cite{Altherr2021, Liu2023b}. Unfortunately, its complexity grows exponentially with $N$, making it intractable for $N  \gtrsim 5$.

For \textit{uncorrelated} noise, one can circumvent this problem by
computing bounds for $\mathcal{F}_\text{AD}$, using the  following iteration~\cite{Kurdzialek2023a}
\begin{equation}
\label{eq:old_it}
\mathcal{F}_\text{AD}^{(l+1)} \le \mathcal{F}_\text{AD}^{(l)}  + 4 \min_{h} \left[     \| \alpha \| +\sqrt{\mathcal{F}_\text{AD}^{(l)}} \| \beta \| \right],
\end{equation}
where  $\beta = \sum_k \dot {\tilde K}_k^\dagger K_k$ (see Appendix~\ref{app:uncorr_der} for a new, simplified derivation).
If there exists $h$ for which $\beta = 0$, $\mathcal{F}_\text{AD}^{(N)}$ scales linearly with $N$ for $N \gg 1$.
Hence
\begin{equation}
\label{eq:uncor_SS}
    \lim_{N \rightarrow \infty } \mathcal{F}_\text{AD}^{(N)}/N \le 4 \min_h \|\alpha\|~~ \text{s.t.}~~\beta=0.
\end{equation}
If instead $\beta \ne 0$ for all $h$, HS is in principle allowed, and the asymptotic form of \eqref{eq:old_it} reads
\begin{equation}
\label{eq:uncor_HS}
    \lim_{N \rightarrow \infty } \mathcal{F}_\text{AD}^{(N)}/N^2 \le 4 \min_h \|\beta\|^2.
\end{equation}
The iterative bound \eqref{eq:old_it} and asymptotic bounds \eqref{eq:uncor_SS}, \eqref{eq:uncor_HS} can be formulated as SDPs~\cite{Koodynski2013, Kurdzialek2023a}.
The bound~\eqref{eq:old_it} is not tight in general, but its asymptotic limits \eqref{eq:uncor_SS} and \eqref{eq:uncor_HS} are always saturable using a parallel strategy~\cite{Zhou2020}.

\paragraph*{Bound for correlated noise.}
In what follows, we present novel bounds analogous to \eqref{eq:old_it}, \eqref{eq:uncor_SS}, \eqref{eq:uncor_HS}  valid for all correlated noise models. Unlike the exact comb QFI formula \eqref{eq:comb_QFI}, these bounds are efficiently calculable for arbitrarily large $N$. 

The comb $\Lambda_\theta^{(N)}$ representing a correlated noise metrological model is a link product of its teeth $\Lmb_\theta : \mathcal{L}(\mathcal{H}_{2i-1} \otimes \mathcal{R}_{i-1}) \rightarrow \mathcal{L}(\mathcal{H}_{2i}\otimes \mathcal{R}_{i}) $ for $i \in \{1,2,...,N\}$, where $\mathcal{R}_i$ are inaccessible environmental spaces modeling correlations; the fixed state $\sigma_\text{in}$ inputs the first register $\mathcal{R}_0$, the last register $\mathcal{R}_N$ is traced out---see Fig.~\ref{fig:corr}(a). 
The $\leftrightarrow$ symbol in  $\Lmb_\theta$ indicates unconcatenated environmental spaces.
Subsequent teeth are assumed identical, though this assumption can be easily dropped.

We divide $\Lambda_\theta^{(N)}$ into  blocks $\Lmb_\theta^{(m)}$ of $m$ teeth each (Fig.~\ref{fig:corr}(b)), intertwined with control combs $\t{C}^{(l,m)}$ connected via ancillae $\mathcal{A}_l$. Any global strategy  $\t{C}^{(N)}$ can be simulated using appropriate $C^{(l,m)}$ with sufficiently large $\mathcal{A}_l$.

Notably, combs $\t{C}^{(l,m)}$ control not only $\mathcal{H}$ and $\mathcal{A}$  at each protocol state, but also environmental subspaces $\mathcal{R}$ between different blocks  $\Lmb_\theta^{(m)}$. In a normal AD scheme, the environment is always directly sent to a next tooth of $\Lambda_\theta^{(N)}$, but here we allow for arbitrary operations on environment every $m$ teeth.
These ``environment leakages'' often make the bound less tight, but are necessary when dividing $N$ correlated channels into blocks of $m$ while keeping the validity of the bound---neglecting the environmental information  may lead to underestimated result.

Additionally, let us replace the joint mixed state $\rho_\theta^{(l)}$ of $\mathcal{H}, \mathcal{R}, \mathcal{A}$ after each block $\Lmb_\theta^{(m)}$ with its QFI NIP $\ket{\Psi_\theta^{(l)}}$.
This  can only increase the final QFI, since any operation possible with a state is also possible with its purification. 

Through purification, the space carrying the information about $\rho_\theta^{(l)}$ and $\dot \rho_\theta^{(l)}$ reduces to the qubit subspace $\mathcal{V}_l= \text{span} (\ket{\Psi_\theta^{(l)}}, \ket{\dot \Psi_\theta^{(l)}})$. After fixing a basis of $\mathcal{V}_l$ and constructing a comb $\mathbb{\Lambda}_\theta^{(l,m)} = \ket{\Psi_\theta^{(l)}} \bra{\Psi_\theta^{(l)}} \otimes \Lmb_\theta^{(m)}$ containing the purified and ``qubitized'' $\rho_\theta^{(l)}$ together with next $m$ teeth, we get (see Appendix~\ref{appD1} for a detailed derivation) $\mathcal{F}_\text{AD}^{(i)} \le F^{(i)}$, where $F^{(0)} = 0$, 
\begin{multline}
\label{eq:new_bound_it}
    F^{(l+m)} = \max_{\t{C}^{(l,m)}} F(\t{C}^{(l,m)} \star \mathbb{\Lambda}_\theta^{(l,m)}), ~~ \mathbb{\Lambda}_\theta^{(l,m)} = \begin{bmatrix}
           1&0 \\
           0 &0\\
         \end{bmatrix} \otimes \varLmb_\theta^{(m)}, \\
         \dot{\mathbb{\Lambda}}^{(l,m)}_\theta = \begin{bmatrix}
           1&0 \\
           0 &0\\
         \end{bmatrix} \otimes \dot {\varLmb}_\theta^{(m)} + \frac{\sqrt{F^{(l)}}}{2} \begin{bmatrix}
           0&1\\
           1 &0\\
         \end{bmatrix} \otimes \varLmb_\theta^{(m)},
\end{multline}
maximizing over $\t{C}^{(l,m)}$ amounts to computing the comb QFI of $\mathbb{\Lambda}_\theta^{(l,m)}$.

The bound generated by \eqref{eq:new_bound_it} can be applied to uncorrelated noise---then additional QFI NIPs are the only reason for the bound untightness, making it at least as tight as  \eqref{eq:old_it}.
The new bound can be then tighter than the old one even for $m=1$, increasing $m$ further tightens \eqref{eq:new_bound_it}, but only for finite $N$---see Appendix~\ref{app:uncorr_ex}.

For correlated noise, for which  \eqref{eq:old_it} does not apply, the role of $m$ is more significant, since information leaks from the environment every $m$ steps. Consequently, increasing $m$ tightens the bound also asymptotically. 

Using \eqref{eq:new_bound_it}, one can prove that  (see Appendix~\ref{appD2}) :
\begin{multline}
\label{eq:itmaxmax}
F^{(l+m)} \le F^{(l)} + 4 \min_h  \left[  a^{(m)} + \sqrt{F^{(l)}}  b^{(m)} \right],   a^{(m)} = \\ =   \max_{\tilde C^{(m)}} \text{Tr} \left( \boldsymbol{\alpha}^{(m)} \tilde C_{00}^{(m)}   \right),~~b^{(m)} =  \max_{\tilde C^{(m)}}   \text{Re}~ \text{Tr} \left(  \boldsymbol{\beta}^{(m)} \tilde C_{10}^{(m)} \right), 
\end{multline}
where $\tilde C^{(m)}$ is a comb of the same type as any $ \tilde C^{(l,m)}= \text{Tr}_{\mathcal{A}_{l+m}} C^{(l,m)}$, $\boldsymbol{\alpha}^{(m)}$ is a performance operator of $\varLmb_\theta^{(m)}$,  $\boldsymbol{\beta}^{(m)} $ is a comb-adapted analogue of $\beta$ (defined as $\boldsymbol{\alpha}^{(m)}$, but without derivative acting on $\bra {\bullet}$ part), $\tilde C^{(m)}_{ij} = \prescript{}{\mathcal{V}_l} {\braket{i|\tilde C^{(m)}|j}}_{\mathcal{V}_l}$. Equation \eqref{eq:itmaxmax} is similar to \eqref{eq:old_it}, but $\| \alpha \|$ and $\| \beta \|$ are replaced with their comb-adapted analogues $a^{(m)}$, $b^{(m)}$.
Consequently, we get asymptotic bounds analogous to \eqref{eq:uncor_SS} and \eqref{eq:uncor_HS} valid for the correlated case; when $b^{(m)} = 0$ for some $h$, then 

\begin{equation}
    \label{eq:cor_SS}
    \lim_{N \rightarrow \infty } \frac{\mathcal{F}_\text{AD}^{(N)}}{N} \le \frac{4}{m} \min_h  a^{(m)} ~ \text{s.t.}~~b^{(m)}=0,
\end{equation}
so the QFI scales at most linearly.
If $b^{(m)} \ne 0$ for all $h$, then HS is allowed:
\begin{equation}
    \label{eq:cor_HS}
    \lim_{N \rightarrow \infty } \frac{\mathcal{F}_\text{AD}^{(N)}}{N^2} \le \frac{4}{m^2} \min_h  {b^{(m)}}^2.
\end{equation}
In Appendix~\ref{app:corr_SDP} we  show how to translate \eqref{eq:new_bound_it}, \eqref{eq:cor_SS} and \eqref{eq:cor_HS} into single SDPs using the technique from Refs.~\cite{Altherr2021,Liu2023b}. Even though bounds \eqref{eq:new_bound_it}, \eqref{eq:cor_SS}, \eqref{eq:cor_HS} are not generally tight, they can be tightened by increasing $m$ (which increases the complexity of the resulting SDPs).

\paragraph*{Example: correlated dephasing.}
To illustrate the practical relevance of the new bounds, we consider phase estimation in the presence of correlated dephasing.
Single-qubit dephasing shrinks the Bloch vector by a factor  $\eta= \cos(\epsilon)$ towards the axis defined by a unit vector $\vec n$.
It corresponds to a random rotation by angle $+ \epsilon$ or $-\epsilon$ around $\vec n$, each with probability $1/2$.

To introduce basic correlations, we assume the consecutive rotational signs $r_i \in \{+,-\}$ follow a binary Markov chain given by conditional probabilities \cite{Kurdzialek2024}
\begin{equation}
    \label{eq:pcond}
    S_{i|i-1}(r_i|r_{i-1}) = (1+r_i r_{i-1} C)/2,
\end{equation}
where $C \in [-1,1]$ controls correlation: $C = 0$ means no correlation, $C = \pm1$ gives maximal (anti-)correlation .
The corresponding quantum model $\Lambda_\theta^{(N)}$ is constructed by alternating unitaries $V_\theta =R_{\hat z}(\theta) R_{\vec n}({+ \epsilon}) \otimes \ket{+}\bra{+}  + R_{\hat z}(\theta) R_{\vec n}({- \epsilon})  \otimes \ket{-}\bra{-} $ acting on $\mathcal{H}$ and $\mathcal{R}$ with mixing operations $\t{S}$ performing stochastic map \eqref{eq:pcond} on basis $\ket{\pm}$ of $\mathcal{R}$  \cite{Kurdzialek2024},  $R_{\vec n} (\varphi) = e^{- \frac{i}{2} \varphi \vec n \cdot \vec \sigma }$, $ \vec \sigma = [\sigma_x, \sigma_y, \sigma_z]$, notice that noise (random rotation by angle $\pm \epsilon$) precedes signal (rotation by $\theta$).

\begin{figure*}[t]
\includegraphics[width=0.95\linewidth]{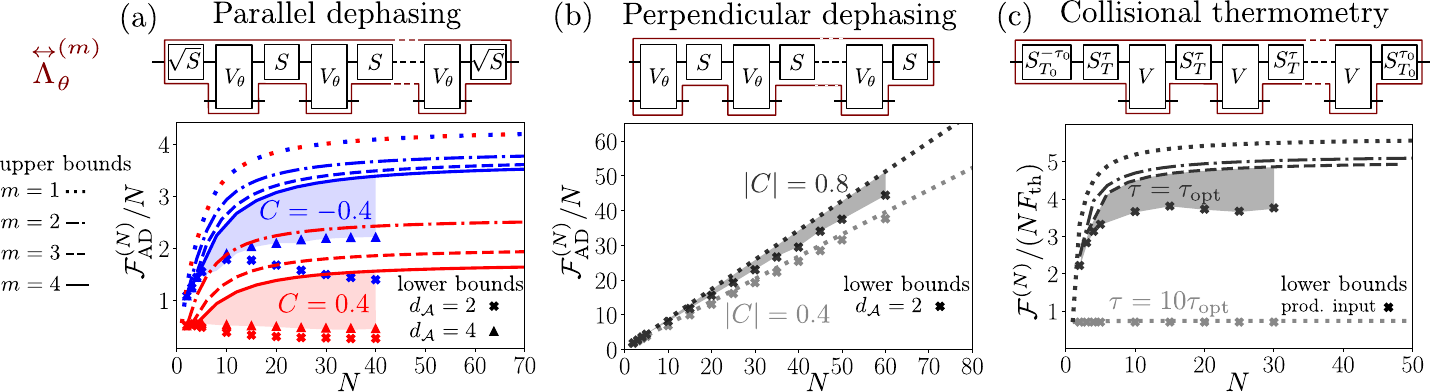}
\caption{ 
Combs $\Lmb_\theta^{(m)}$ and resulting  upper bounds for QFI for three different models. Lower bounds obtained using tensor networks techniques \cite{Kurdzialek2024, Chabuda2020} are shown for comparison, areas between the corresponding tightest lower  and upper bounds are shaded. 
}
\label{fig:deph}
\end{figure*}

%
The comb $\Lambda_\theta^{(N)}$ can be split into pieces $\Lmb_\theta^{(m)}$ in different ways, and this choice affects the tightness of the resulting bound.
First, we consider \textit{parallel dephasing} $\vec n = \hat z$, when noise commutes with signal.
If we cut the chain after $\t{S}$ and before $V_\theta$, we get information on the sign of $\epsilon$ for the first dephasing channel in each block, while cutting the chain after $V_\theta$ and before $\t{S}$, we get this information for the \textit{last} channel in each block.
Consequently, the first/last channel in each block is effectively noiseless, so the bound manifests an unrealistic HS.

To resolve this issue, we write $\t S$ as a concatenation of two identical stochastic maps, $\t S = \sqrt{\t S} \circ \sqrt{\t S} $, see Appendix~\ref{appE1} for exact construction of $\sqrt{\t S}$.
The chain cuts are then made in the middle of every $m$-th map $\t S$, the resulting  $ \Lmb_\theta^{(m)}$ consists of  $V_\theta$  with $\t S$ between them and $\sqrt{\t S}$  at both ends, see Fig.~\ref{fig:deph}(a).
We insert $ \Lmb_\theta^{(m)}$ into the recursive procedure \eqref{eq:new_bound_it} to derive bounds,
sending the maximally mixed state as input for $\mathcal{R}$ for the first iteration (probabilities of $\pm$ are equal in the first channel).

We observe that the HS is not possible for correlated parallel dephasing for any $-1<C<1$ and $0 \le \eta <1$.
In Fig.~\ref{fig:deph}(a) we compare upper bounds, calculated for positive ($C=0.4$) and negative ($C=-0.4$) correlations for $\eta = 0.75$, with lower bounds---QFIs achievable using protocols with small ancilla dimension $d_\mathcal{A}$, calculated using the tensor-network algorithm from Refs.~\cite{Kurdzialek2024, dulian_package} .
These bounds can be made arbitrarily tight by increasing $m$ and $d_\mathcal{A}$, respectively; however, this quickly becomes numerically costly.
We performed
calculations up to $d_\mathcal{A} = 4$ and $m=4$, for which the lower and upper bounds still do not coincide.
Nevertheless, this is the first time that  
precision limits of metrological protocols for correlated noise with arbitrarily large $d_\mathcal{A}$ are explicitly evaluated.
This allows us to deduce that negative correlations offer metrological advantage over positive correlations.

We also study \emph{perpendicular dephasing} $\vec n = \hat x$, for which the HS is possible~\cite{Kurdzialek2023a}.
In this case, we observed that the tightest bound is obtained
by splitting the comb after $V_\theta$ and before $\t S$, see Fig.~\ref{fig:deph}(b).
This is because non-commuting noise acts before the signal, and the information about rotation sign is useless after the signal.
Interestingly, the resulting bounds are equally tight for $m \in \{1,2,3\}$, suggesting that the bound is already tight for $m=1$.
This is further supported by the observation that
the QFI achievable by an adaptive protocol with $d_\mathcal{A} = 2$ is very close to the upper bound calculated for $m=1$, see Fig.~\ref{fig:deph}(b).
We also show that for perpendicular dephasing the QFI does not depend on the sign of $C$---all correlations act as an extra source of information, since the achievable precision increases with $|C|$.

\paragraph*{Example: collisional thermometry.}
From a different perspective, the environment register $\mathcal{R}$ may not represent noise, but an inaccessible quantum system, measured indirectly by sequential interactions with the mediators $\mathcal{H}_k$.
This setup corresponds to quantum collision models~\cite{Ciccarello2022} and, for factorized probes, to quantum Markov chains, widely studied for parameter estimation~\cite{Guta2011,Godley2023}.

We consider a simple realization of \textit{collisional quantum thermometry}~\cite{Seah2019, Shu2020,OConnor2021,Mendonca2024}, where a thermometer qubit ($\mathcal{R}$) probes a thermal bath, whose temperature $T$ is to be estimated by measuring mediator qubits ($\mathcal{H}$) that sequentially interact with the thermometer.
Each collision is a partial energy exchange between $\mathcal{H}$ and $\mathcal{R}$ governed by the unitary $V = e^{-i g t (\sigma_+ \otimes \sigma_- + \sigma_- \otimes \sigma_+)}$, $g$ is a coupling constant, $t$ is an interaction time, $\sigma_\pm = (\sigma_x \pm i \sigma_y)$; a full energy swap occurs for $gt = 0.5 \pi$.  Between collisions, the thermometer evolves via a non-unitary channel  $\t{S}_T^\tau (\rho) = e^{\tau \mathcal{L}_T} (\rho)$, where $\mathcal{L}_T$ is a thermalizing Lindbladian, see \cite{Seah2019} and Appendix~\ref{appE2} for an exact definition. For $\tau \rightarrow \infty$, any input state gets fully thermalized, $\t{S}_T^\infty (\rho) = \rho_{\t{th}, T} $, where $\rho_{\t{th}, T} = e^{-H_\mathcal{E} / k_B T} / Z$ is a Gibbs state, $H_\mathcal{E} = \hbar \Omega \sigma_z /2$. We assume that the thermometer is initialized in a thermal state $\rho_{\t{th}, T}$. \footnote{This assumption differs from Ref.~\cite{Seah2019, Shu2020}, in which the initial state was assumed to be stationary state of both $\t{S}_T^\tau$ and $V$. However, this assumption does not affect the asymptoptotic results. }

Various strategies have been explored in this setting \cite{Seah2019, Shu2020}, yet their rigorous study for large $N$ 
is challenging due to environment-induced correlations among measured mediator qubits. Our bounds fully account for these correlations.
Since parallel schemes are a subset of adaptive ones, the bounds also apply to non-adaptive scenarios with arbitrary initial entanglement among mediators and a final collective measurement.

We define \textit{thermal Fisher information} as $F_\t{th} = F( \rho_{\t{th}, T})$.
Interestingly, the channel QFI of $\t{S}_T^\tau$ can exceed $F_\t{th}$ for finite $\tau$, reaching $\mathcal{F}(\t{S}_T^{\tau_\t{opt}}) \approx 5.66 F_{\t{th}, T}$ at the optimal thermalization time $\tau_\t{opt}$ when $k_B T / \hbar \Omega = 2$ \cite{Seah2019}.
We consider a collisional model with $gt = 0.35 \pi < 0.5 \pi$, which leads to correlations among output mediators. We then compute upper bounds for $\mathcal{F}_\t{AD}$ for $m = 1,2,3$ in two regimes: $\tau = \tau_\t{opt}$ and $\tau = 10 \tau_\t{opt}$. 

To tighten the bounds, we insert an additional thermalization map \(\t{S}_{T_0}^{\tau_0}\) at the end of each \(\Lmb_\theta^{(m)}\), where \(T_0\) denotes a fixed, reference temperature---i.e., \(\frac{d}{dT} \t{S}_{T_0}^{\tau_0} = 0\), so the map is insensitive to \(T\), see Fig.~\ref{fig:deph}(c). To preserve the overall channel structure, we prepend an inverse map \(\t{S}_{T_0}^{-\tau_0}\), ensuring the two cancel out when constructing \(\Lambda_\theta^{(N)}\) from the building blocks \(\Lmb_\theta^{(m)}\). We choose \(\tau_0 < \tau\) to keep \(\t{S}_T^\tau \circ \t{S}_{T_0}^{\tau_0}\) completely positive, and optimize \(\tau_0\) to obtain the tightest possible bounds.

For $\tau = \tau_{\t{opt}}$, correlations play a significant role, as reflected by the tightening of asymptoptotic bounds with increasing $m$ (see Fig.~\ref{fig:deph}(c)). In contrast, for $\tau = 10 \tau_{\t{opt}}$, the output of $\t{S}_T^\tau$ is nearly thermal for any input, correlations are negligible, and the bounds are almost insensitive to $m$.

To assess tightness, we compare our upper bounds with lower bounds obtained by optimizing the output QFI over input product states. This optimization, performed efficiently via tensor networks~\cite{Chabuda2020, dulian_package}, shows near saturation of the upper bound for $\tau = 10 \tau_{\t{opt}}$, confirming that product states are optimal in this regime. For $\tau = \tau_{\t{opt}}$, a gap remains, but the bounds still closely follow the QFI trend, indicating limited potential gain from entanglement or adaptivity.

%

\paragraph*{Conclusions.} 
These examples represent only a fraction of the possible applications of the new bounds.
 The formalism allows one to handle both classical and inherently quantum correlations, i.e. generic non-Markovian open quantum systems \cite{Caruso2014,deVega2017,Pollock2018a,Milz2021}.
 Although the sequential scheme suggests a focus on temporal correlations, the bound is also valid for analogous spatial correlations, where channels are sampled in parallel, since this is a subset of adaptive strategies.
Furthermore, this approach provides a tighter bound for uncorrelated noise than the current state of the art \cite{Kurdzialek2023a}. 

This work is complementary to Refs.~\cite{Kurdzialek2024, Liu2024}, where a tensor network approach was proposed to find AD estimation strategies with a limited ancilla dimension, providing a \textit{lower} bound on $\mathcal{F}_\text{AD}$.
In combination, these works establish a comprehensive framework to study metrological protocols in the presence of correlated noise, for a large number of probed channels.
The package containing the functions to compute the bounds as well as perform tensor-network optimization to obtain the lower bounds as in Fig.~\ref{fig:deph} is now publicly available \cite{dulian_package}.

A challenging extension will be to proceed towards the continuous time limit, where quantum combs (known as process tensors~\cite{Milz2021}) are routinely used in numerical studies of the dynamics and control of non-Markovian open systems~\cite{Pollock2018a,Jorgensen2019,Fux2021, Butler2024,Ortega-Taberner2024}.
Finally, it will be interesting to see if similar ideas apply to the related problem of discriminating quantum combs~\cite{Chiribella2008a,Gutoski2012,Wang2019e,Nakahira2021,Nakahira2021b,Hirche2023b,Zambon2024a,Bavaresco2021}.

\paragraph*{Acknowledgements.} 
We thank Wojciech Górecki and Andrea Smirne for helpful discussions.
This work was supported by the National Science Center (Poland) grant No.2020/37/B/ST2/02134. 
FA acknowledges support from Marie Skłodowska-Curie Action EUHORIZON-MSCA-2021PF-01 (project \mbox{QECANM}, grant No. 101068347).

\bibliography{PRL_main_only_v1}
\onecolumngrid
\appendix

\onecolumngrid
\appendix  

\section{Basics of Quantum Fisher Information}
\label{app:QFI}
The quantum Fisher information  (QFI) of an arbitrary mixed state $\rho_\theta$ can be calculated using the formula
\begin{equation}
\label{eq:app_QFI_def}
F(\rho_\theta) = \t{Tr}(\rho_\theta L_\theta^2), \quad \dot \rho_\theta = \frac{1}{2} \left( \rho_\theta L_\theta + L_\theta \rho_\theta \right),
\end{equation}
where dot denotes the derivative over $\theta$, and the symmetric logarithmic derivative (SLD) matrix $L_\theta$ can be calculated by solving the rightmost equation from \eqref{eq:app_QFI_def}. For pure state models $\rho_\theta = \ket{\psi_\theta} \bra{\psi_\theta}$, the closed analytical formula for the SLD and the QFI can be easily derived:
\begin{equation}
\label{eq:app_QFI_pure}
L_\theta = 2 \left( \ket{\dot \psi_\theta} \bra{\psi_\theta} + \ket{\psi_\theta} \bra{\dot \psi_\theta} \right), \quad F(\ket{\psi_\theta}) = 4(\braket{\dot \psi_\theta|\dot \psi_\theta} - |\braket{\psi_\theta| \dot \psi_\theta}|^2).
\end{equation}
When $\ket{\Psi_\theta}$ is a purification of $\rho_\theta$, then $F(\ket{\Psi_\theta}) \ge F(\rho_\theta)$. Additionally, for each $\rho_\theta$ we can construct its QFI non-increasing purification (QFI NIP) satisfying
\begin{equation}
\label{eq:app_QFINIP}
    F(\ket{\Psi_\theta}) = F(\rho_\theta),\quad \braket{\Psi_\theta|\dot \Psi_\theta} = 0
\end{equation}
The second condition can be always satisfied by multiplying $\ket{\Psi_\theta}$ with a global, $\theta$-dependent phase.

\section{Basics of quantum combs}
\label{appB}
Let us introduce the formal definition of quantum combs \cite{Chiribella2008b}. We use roman font for channels and italics for corresponding Choi-Jamiołkowski (CJ) operators: the CJ operator of a channel $\Lambda: \mathcal{H}_\text{in} \rightarrow \mathcal{H}_\text{out}$ is defined as $\varLambda = \Lambda \otimes \mathcal{I} (\ket{\Psi_+} \bra{\Psi_+})$, where $\ket{\Psi_+} = \sum_i \ket{i} \ket{i}$ , $\ket{i}$ is an orthonormal basis of $\mathcal{H}_\text{in}$, $\mathcal{I}$ is the identity channel.
Let $\t{E}$ be a quantum channel with with inputs $\mathcal{K}_1, \mathcal{K}_3, ..., \mathcal{K}_{2N-1}$ and outputs $\mathcal{K}_2, \mathcal{K}_4, ... \mathcal{K}_{2N}$. Then, $\t{E} \in \t{Comb}[(\mathcal{K}_1, \mathcal{K}_2), (\mathcal{K}_3, \mathcal{K}_4),...,(\mathcal{K}_{2N-1}, \mathcal{K}_{2N})]$ if and only if (iff) there exists a sequence of operators $ E^{(k)}$ for $k = 1,2,...,N$ such that $E = E^{(N)}$ and:
\begin{enumerate}
    \item $\t{Tr}_{2k} E^{(k)} = E^{(k-1)} \otimes \mathbb{1}_{2k-1}$ for $k=2,3,...,N$
    \item $\text{Tr}_2 E^{(1)} = \mathbb{1}_1$
    \item $E \succeq 0$
\end{enumerate}
Intuitively, $\mathcal{K}_{2k-1}, \mathcal{K}_{2k}$ are input/output spaces of the $k$th tooth of a comb.
The above conditions imply the causal order of the teeth---the output $\mathcal{K}_{2l}$ can depend on the input $\mathcal{K}_{2j-1}$ only if $l \ge j$.
In other words, quantum combs can be used to simulate channels with memory, when each output may depend on all preceding inputs.
Therefore, they are the ideal tool to represent sequences of correlated channels or adaptive quantum strategies.
Notably, a sequence $\t{E}^{(k)}$ consists of quantum combs with increasing number of teeth: $\t{E}^{(k)} \in \t{Comb}[(\mathcal{K}_1, \mathcal{K}_2),... (\mathcal{K}_{2k-1}, \mathcal{K}_{2k})]$.
Two combs can be connected with each other using a \textit{link product} operation $\star$ \cite{Chiribella2008b}.
When $\t{E}$, $\t{F}$ are two quantum combs, then $G = E \star F$ is a comb created by connecting the input spaces of $\t{E}$ with the corresponding output spaces of $\t{F}$ and the output spaces of $\t{E}$ with the corresponding input spaces of $\t{F}$ (only common spaces of $\t E$ and $\t F$ get connected).
Let $\t E$, $\t F$ act on sets of spaces $M$, $N$ respectively. The link product is then formally defined using CJ operators :
\begin{equation}
    \t{G} = \t{E} \star \t{F} \iff G = \t{Tr}_{M \cap N} \left[ \left( E^{\t{T}_{M \cap N}} \otimes \mathbb{1}_{N \setminus M} \right) \left( \mathbb{1}_{M \setminus N} \otimes F \right) \right],
\end{equation}
where $M \cap N$ is the set of common (linked) spaces, $\t{T}_{M \cap N}$ is a partial transposition (with respect to common spaces only).
When $M=N$, then $\t{E} \star \t{F}$ is a scalar, using the comb conditions and the fact that each output of $\t{E}$ is an input of $\t{F}$ and \textit{vice versa}, it can be then shown that $\t{E} \star \t{F} = 1$.
For more information about quantum combs and link product see Refs.~\cite{Chiribella2008a, Chiribella2008b, Chiribella2009}.   
\section{Uncorrelated noise bound}
\label{app:uncorr}
The bound \eqref{eq:old_it} has been derived in Ref. \cite{Kurdzialek2023a}, and the asymptotic bounds \eqref{eq:uncor_SS} and \eqref{eq:uncor_HS} are its direct consequences, see equations (11), (12) from Ref. \cite{Kurdzialek2023a}. In what follows, we provide a simpler derivation of \eqref{eq:old_it}. From the new derivation it is clear, that the bound generated by \eqref{eq:new_bound_it} for uncorrelated noise is at least as tight as the one given by \eqref{eq:old_it}. The advantage of the old derivation is that it can be easily generalized to strategies involving causal superpositions.
\subsection{Simplified derivation}
\label{app:uncorr_der}
\begin{figure}[h]
\includegraphics[width=0.45\columnwidth]{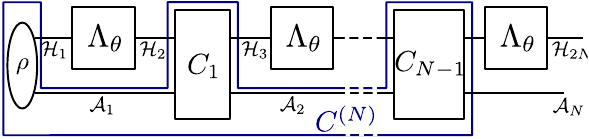}
\caption{Adaptive scheme for uncorrelated channels}
\label{fig:app_uncorr_scheme}
\end{figure}

We consider an adaptive (AD) scheme in which $N$ independent channels $\Lambda_\theta$ are probed (see Fig.~\ref{fig:app_uncorr_scheme}) and keep the notation from the main text.   
 Let $\rho_\theta^{(l)} \in \mathcal{L}(\mathcal{H}_{2l} \otimes \mathcal{A}_l) $ be the probe and ancilla state after $l$th use of $\Lambda_\theta$, and let $F(\rho_\theta^{(l)}) = F^{(l)}$. The state after the next control is $\text{C}_l(\rho_\theta^{(l)})$, we construct its QFI NIP $\ket{\Psi_\theta^{(l)}} \in \mathcal{H}_{2l+1} \otimes \mathcal{A}_{l+1} \otimes \mathcal{E}_l$, where $\mathcal{E}_l$ is an artificial space added for purification purposes. Using \eqref{eq:app_QFINIP} and \eqref{eq:app_QFI_pure} and the fact that a $\theta$-independent channel cannot increase the QFI \cite[p.~57]{Kolodynski2014}, we obtain
\begin{equation}
\label{eq:app_b1}
     4 \braket{\dot \Psi_\theta^{(l)} | \dot \Psi_\theta^{(l)}} = F(\text{C}_l(\rho_\theta^{(l)})) \le F^{(l)}.
\end{equation}
 Let $\bar K_{k, \theta} =  K _{k, \theta} \otimes \mathbb{1}_{\mathcal{A}_{l+1} \otimes \mathcal{E}_l} $, the state of probe and ancilla after the action of the $(l+1)$th channel
 can be written as
\begin{equation}
    \rho_\theta^{(l+1)} = \textrm{Tr}_{\mathcal{E}_l} \left( \sum_k \bar K_{k,\theta} \ket{\Psi_\theta^{(l)}} \bra{\Psi_\theta^{(l)}} \bar K_{k,\theta}^\dagger  \right).
\end{equation}
Since the QFI of a subsystem is smaller or equal than the QFI of the whole system (i) and the QFI of a purification is larger or equal than the QFI of a purified state (ii), we obtain
\begin{equation}
\label{eq:app_QFI_chain_1}
    F^{(l+1)} \overset{(\t{i})}{\le} F \left(\sum_k \bar K_{k,\theta} \ket{\Psi_\theta^{(l)}} \bra{\Psi_\theta^{(l)}} \bar K_{k,\theta}^\dagger \right) \overset{(\t{ii})}{\le} F \left(\sum_k \bar K_{k,\theta} \ket{\Psi_\theta^{(l)}} \otimes \ket{k}_{\mathcal{E'}_l} \right),
\end{equation}
where $\mathcal{E'}_l$ is an additional space added for purification, and the vectors $\ket{k}_{\mathcal{E'}_l}$ form its o.-n. basis.
After expanding the rightmost part of \eqref{eq:app_QFI_chain_1} using \eqref{eq:app_QFI_pure}, we obtain
\begin{align}
\label{eq:app_b2}
F^{(l+1)}/4 &\le \bra{\dot \Psi_\theta^{(l)}} \sum_k \bar K^\dagger_{\theta, k} \bar K_{\theta, k} \ket{\dot \Psi_{\theta}^{(l)}} + \bra{\dot \Psi_\theta^{(l)}} \sum_k \bar K^\dagger_{\theta, k} \dot {\bar K}_{\theta, k} \ket{ \Psi_{\theta}^{(l)}} + \bra{ \Psi_\theta^{(l)}} \sum_k \dot {\bar K}^\dagger_{\theta, k} {\bar K}_{\theta, k} \ket{\dot \Psi_{\theta}^{(l)}} + \\ &+ \bra{ \Psi_\theta^{(l)}} \sum_k \dot {\bar K}^\dagger_{\theta, k} \dot {\bar K}_{\theta, k} \ket{ \Psi_{\theta}^{(l)}} 
\end{align}
From \eqref{eq:app_b1} and the identity $ \sum_k \bar K^\dagger_{\theta, k} \bar K_{\theta, k} = \mathbb{1}$ we deduce that the first term is upper bounded by $F^{(l)}/4$; the last term is upper-bounded by $\| \alpha \|$ because $\braket{ \Psi_\theta^{(l)}| \Psi_\theta^{(l)}} = 1$ and $\| A \otimes \mathbb{1} \| = \| A \|$.
Both the second and the third term are upper-bounded by $\sqrt{F^{(l)}/4} \|\beta \|$, this follows from the inequality
\begin{equation}
    \braket{x|A|y} \le \sqrt{\braket{x|x}} \|A\| \sqrt{\braket{y|y}},
\end{equation}
which is a less general version of (7) from \cite{Kurdzialek2023a}. After taking this all together, we obtain
\begin{equation}
\label{eq:app_old_it}
\mathcal{F}_\text{AD}^{(l+1)} \le \mathcal{F}_\text{AD}^{(l)}  + 4 \min_{h} \left[     \| \alpha \| +\sqrt{\mathcal{F}_\text{AD}^{(l)}} \| \beta \| \right],
\end{equation}
which is \eqref{eq:old_it} from the main text. 

To derive this bound, we assumed to have access to the QFI NIP of probe and ancilla after each control.
Moreover, we maximized each term of \eqref{eq:app_b2} independently---this is another factor making the bound not tight, since usually all the terms cannot be maximal at the same time. Interestingly, to derive \eqref{eq:new_bound_it}, we only assumed  to have access to the QFI NIP every $m$ steps, no additional assumptions were required (see the exact derivation in the next Section).
Therefore, the bound generated by~\eqref{eq:new_bound_it} is guaranteed to be at least as tight as the one generated by \eqref{eq:old_it} already for $m=1$.
Increasing $m$ may tighten the bound even more.

\subsection{Example}
\label{app:uncorr_ex}
\begin{figure}[H]
\centering
\includegraphics[width=0.45\columnwidth]{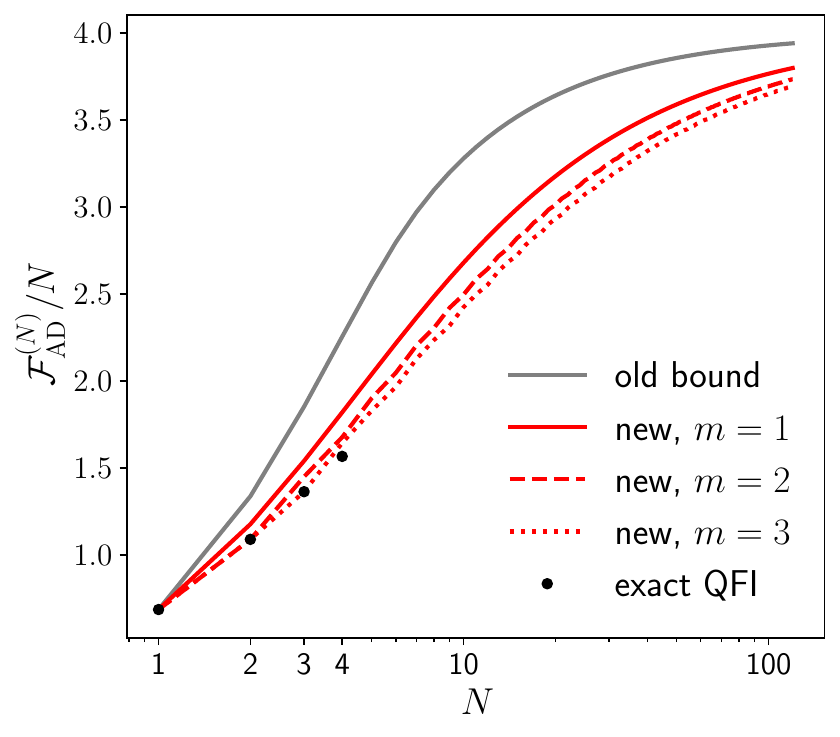}
\caption{Precision bounds for phase estimation in the presence of amplitude damping noise. }
\label{fig:app_damp}
\end{figure}

To illustrate the effectiveness of the new bound for uncorrelated noise, let us calculate it for $m \in \{1,2,3\}$ for phase estimation in the presence of amplitude damping noise, for which $K_{k,\theta} = K_k e^{-\frac{i \theta \sigma_z}{2}}$, with
\begin{equation}
    K_1 = \ket{0} \bra{0} + \sqrt{p} \ket{1} \bra{1}, \quad K_2 = \sqrt{1-p} \ket{0} \bra{1}.
\end{equation}
In Fig.~\ref{fig:app_damp}, we demonstrate the results for $p=0.5$. The newly introduced bound is tighter than the old one already for $m=1$ . For larger $m$, the bound becomes even tighter, and very close to the optimal QFI calculated exactly for $m \le 4$  using the algorithm from Ref. \cite{Altherr2021}.
\section{Correlated noise bounds}
\label{app:corr}
In this Section, we provide a detailed derivation of  the recursive bounds \eqref{eq:new_bound_it} and \eqref{eq:itmaxmax} valid for correlated noise. 
Then, we derive the asymptotic bounds \eqref{eq:cor_SS} and \eqref{eq:cor_HS} using \eqref{eq:itmaxmax}.
Finally, we demonstrate how to formulate these new recursive and asymptotic bounds as SDPs. 

\subsection{Derivation of recursive bound \eqref{eq:new_bound_it} }
\label{appD1}
\begin{figure}[H]
\centering
\includegraphics[width=0.45\columnwidth]{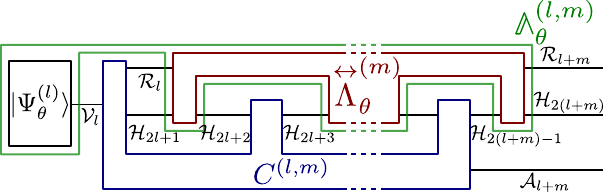}
\caption{The scheme of comb $\mathbb{\Lambda}_\theta^{(l,m)}$ and construction of the output state $\rho_\theta^{(l+m)}$}
\label{fig:app_Lcomb}
\end{figure}

Consider the joint state of the system, ancilla, and environment after $l$ teeth of the comb $\Lambda_\theta^{(N)}$:  $\rho_\theta^{(l)} \in \mathcal{H}_{2l} \otimes \mathcal{R}_l \otimes \mathcal{A}_l$, see Fig.~1 in the main text.
To upper bound $F(\rho_\theta^{(l+m)})$ in terms of $F(\rho_\theta^{(l)})$, we replace $\rho_\theta^{(l)}$ with its QFI NIP $\ket{\Psi_\theta^{(l)}}$. This substitution can only increase the resulting QFI, since any operation possible on the original state is also possible on its purification. Next, we derive an upper bound on the QFI of the state obtained by evolving  $\ket{\Psi_\theta^{(l)}}$ through the next $m$ teeth, represented by $\Lmb_\theta^{(m)}$.
Arbitrary adaptive control is allowed during this evolution. To calculate this upper bound, let us construct a comb $\mathbb{\Lambda}_\theta^{(l,m)} = \ket{\Psi_\theta^{(l)}} \bra{\Psi_\theta^{(l)}} \otimes \varLmb_\theta^{(m)} $, such that $\ket{\Psi_\theta^{(l)}} \bra{\Psi_\theta^{(l)}}$ is its first tooth (with trivial input), see Fig.~~\ref{fig:app_Lcomb}.

Then, the total output state of $\mathcal{H}_{2(l+m)} \otimes \mathcal{R}_{l+m} \otimes \mathcal{A}_{l+m}$ can be written as   $\t C^{(l,m)} \star \mathbb{\Lambda}_\theta^{(l,m)}$, where $\t C^{(l,m)}$ is an arbitrary comb representing adaptive control, $\t C^{(l,m)}$ acts on the spaces depicted in Fig.~\ref{fig:app_Lcomb}. Notice, that $C^{(l,m)}$ acts also on environmental subspace $\mathcal{R}_l$, which is impossible in a normal AD scheme. Therefore, the QFI of $\t C^{(l,m)} \star \mathbb{\Lambda}_\theta^{(l,m)}$ might be larger than $F (\rho_\theta^{(l+m)})$, moreover $F(\rho_\theta^{(l+m)}) \le  \max_{\t C^{(l,m)}} F(\t C^{(l,m)} \star \mathbb{\Lambda}_\theta^{(l,m)})$ because all types of controls allowed in a normal AD scheme can be simulated by a proper choice of $\t C^{(l,m)}$. 
These ``environmental leakages'' that happen every $m$ teeth are one of the reasons for the potential looseness of the bound.

We are now nearly ready to compute the upper bound for $F(\rho_\theta^{(l+m)})$,  expressed as the comb QFI of  $\mathbb{\Lambda}_\theta^{(l,m)}$, which can be evaluated for any comb using the results of Refs.~\cite{Altherr2021, Liu2023b}. To fully specify $\mathbb{\Lambda}_\theta^{(l,m)}$, we define the basis of $\mathcal{V}_l = \text{span}(\ket{\Psi_\theta^{(l)}}, \ket{\dot \Psi_\theta^{(l)}} )$:
\begin{equation}
\label{eq:app_01basis}
    \ket{0} =\ket{\Psi_\theta^{(l)}} , \quad \ket{1} = 2 \ket{\dot \Psi_\theta^{(l)}} / \sqrt{F(\rho_\theta^{(l)})}. 
\end{equation}
 This basis is orthonormal, which follows directly from quantum state normalization and from properties of the QFI NIP~\eqref{eq:app_QFINIP}.
 The density matrices of the state and its derivative can be expressed in this basis as
 \begin{equation}
     \ket{\Psi_\theta^{(l)}} \bra{\Psi_\theta^{(l)}} = \begin{bmatrix}
           1&0 \\
           0 &0\\
         \end{bmatrix} , \quad \frac{\t{d}}{\t{d} \theta} \left(   \ket{\Psi_\theta^{(l)}} \bra{\Psi_\theta^{(l)}}\right) = \frac{\sqrt{F(\rho_\theta^{(l)})}}{2} \begin{bmatrix}
           0&1\\
           1 &0\\
         \end{bmatrix}
 \end{equation}

Finally, we can explicitly write the bound for $F(\rho_\theta^{(l+m)})$ as $F(\rho_\theta^{(l+m)}) \le F^{(l+m)}$, where $F^{(km)}$ is defined by the following iteration
\begin{equation}
\label{eq:app_itnew}
   F^{(0)} = 0, \quad  F^{(l+m)} = \max_{C^{(l,m)}} F(C^{(l,m)} \star \mathbb{\Lambda}_\theta^{(l,m)}),~~\mathbb{\Lambda}_\theta^{(l,m)} =\begin{bmatrix}
           1&0 \\
           0 &0\\
         \end{bmatrix} \otimes \varLmb_\theta^{(m)},~~\dot{\mathbb{\Lambda}}^{(l,m)}_\theta = \begin{bmatrix}
           1&0 \\
           0 &0\\
         \end{bmatrix} \otimes \dot {\varLmb}_\theta^{(m)} + \frac{\sqrt{F^{(l)}}}{2} \begin{bmatrix}
           0&1\\
           1 &0\\
         \end{bmatrix} \otimes \varLmb_\theta^{(m)},
\end{equation}
where $F(\rho_\theta^{(l)})$ was changed to its upper bound $F^{(l)}$ in the expression for $\frac{\t{d}}{\t{d} \theta} \left(   \ket{\Psi_\theta^{(l)}} \bra{\Psi_\theta^{(l)}}\right)$. 

The maximization over $\t C^{(l,m)}$ can be performed using the SDP described in Refs.~\cite{Altherr2021, Liu2023b}.
The iteration \ref{eq:app_itnew} gives an upper bound for $F(\rho_\theta^{(l)})$ valid for all control combs $\t C$, so it is also an upper bound for $\mathcal{F}_\text{AD}^{(l)}$.
Interestingly, for uncorrelated cases there are no ``environmental leakages'', so the only reason for the lack of tightness of the bound is that the state $\rho_\theta^{(l)}$ is replaced with its QFI NIP every $m$ steps.
This makes the bound \eqref{eq:app_itnew} at least as tight as \eqref{eq:app_old_it} for uncorrelated cases. 

\subsection{Derivation of recursive bound \eqref{eq:itmaxmax}}
\label{appD2}
 Let $\varLmb_\theta^{(m)} = \sum_k \ket{K_{k,\theta}^{(m)}} \bra{K_{k,\theta}^{(m)}}$ , since $\mathbb{\Lambda}_\theta^{(l,m)} = \ket{\Psi_\theta^{(l)}} \bra{\Psi_\theta^{(l)}} \otimes \varLmb_\theta^{(m)}$, we can decompose
\begin{equation}
\label{eq:app_Lmb_decomp}
    \mathbb{\Lambda}_\theta^{(l,m)} =\sum_k \ket{L_{k,\theta}^{(l,m)}} \bra{L_{k,\theta}^{(l,m)}},
\end{equation}
 where $\ket{L_{k,\theta}^{(l,m)}} = \ket{\Psi_\theta^{(l)}} \otimes \ket{K_{k,\theta}^{(m)}}$. Using Leibniz rule and \eqref{eq:app_01basis} we obtain 
 \begin{equation}
 \label{eq:app_Lmb_decomp2}
     \ket{ { L}_{k,\theta}^{(l,m)}} = \ket{0} \otimes \ket{ { K}_{k,\theta}^{(m)}}, \quad \ket{\dot { L}_{k,\theta}^{(l,m)}} = \ket{0} \otimes \ket{\dot { K}_{k,\theta}^{(m)}} + \sqrt{F^{(l)}}/2 \ket{1} \otimes \ket{ K_{k,\theta}^{(m)}}.
 \end{equation}
Let $\ket{\dot {\tilde K}_{k,\theta}^{(m)}} = \ket{\dot { K}_{k,\theta}^{(m)}} - i h_{k k'} \ket{ { K}_{k',\theta}^{(m)}}$, where $h$ is a hermitian matrix generating equivalent Kraus representations of a channel; the summation is performed over repeated indices. Then, $\ket{\dot {\tilde L}_{k,\theta}^{(l,m)}} = \ket{0} \otimes \ket{\dot {\tilde K}_{k,\theta}^{(m)}} + \sqrt{F^{(l)}}/2 \ket{1} \otimes \ket{ K_{k,\theta}^{(m)}}$, where $\ket{\dot {\tilde L}_{k,\theta}^{(l,m)}} = \ket{\dot { L}_{k,\theta}^{(l,m)}} - i h_{k k'} \ket{ { L}_{k',\theta}^{(l,m)}}$. The performance operator of $\Lmb_\theta^{(m)}$ is $\boldsymbol{\alpha}^{(m)} = \text{Tr}_\text{out} \left( \sum_k \ket{\dot {\tilde K}_{k,\theta}^{(m)}} \bra{\dot {\tilde K}_{k,\theta}^{(m)}} \right)$, the performance operator of $\mathbb{\Lambda}_\theta^{(l,m)}$ is $\boldsymbol{\alpha}_\mathbb{\Lambda} ^{(l,m)} = \text{Tr}_\text{out} \left( \sum_k \ket{\dot {\tilde L}_{k,\theta}^{(l,m)}} \bra{\dot {\tilde L}_{k,\theta}^{(l,m)}} \right)$, here $\text{Tr}_\text{out}$ is the partial trace over the subspace $\mathcal{H}_{2(l+m)}\otimes \mathcal{R}_{l+m}$, which is the last output of  $\Lmb_\theta^{(m)}$ and of $\mathbb{\Lambda}_\theta^{(l,m)}$.  Using \eqref{eq:app_Lmb_decomp2}, we get :
\begin{equation}
\label{eq:app_perf_op}
     \boldsymbol{\alpha}^{(l,m)}_\mathbb{\Lambda} = \ket{0}\bra{0} \otimes \boldsymbol{\alpha}^{(m)} + \frac{1}{2}\sqrt{F^{(l)}} \ket{0} \bra{1} \otimes \boldsymbol{\beta}^{(m)} + \frac{1}{2}\sqrt{F^{(l)}} \ket{1} \bra{0} \otimes \boldsymbol{\beta}^{ (m) \dagger} + \frac{1}{4} F^{(l)} \ket{1} \bra{1} \otimes \tilde \varLambda^{(m)},
\end{equation}
where $\tilde \varLambda^{(m)} = \text{Tr}_\text{out} \varLmb_\theta^{(m)}$,  $\boldsymbol{\beta}^{(m)} = \text{Tr}_\text{out} \left( \sum_k \ket{\dot {\tilde K}_{k,\theta}^{(m)}} \bra{ K_{k,\theta}^{(m)}} \right)$.
After writing the maximization in \eqref{eq:app_itnew} using the expression for the comb QFI \eqref{eq:comb_QFI} in terms of the performance operator, we get
\begin{equation}
    F^{(l+m)} = \min_h \max_{\tilde C^{(m,l)}} 4 \text{Tr} \left( \boldsymbol{\alpha}_\mathbb{\Lambda} ^{(l,m)} \tilde C^{(l,m)} \right),
\end{equation}
where $\tilde C^{(l,m)} = \text{Tr}_{\mathcal{A}_{l+m}} C^{(l,m)}$. 
After decomposing $\tilde C^{(l,m)} = \sum_{i,j=0}^1 \ket{i}_{\mathcal{V}_l} \bra{j} \otimes \tilde C_{ij}^{(l,m)}$, using \eqref{eq:app_perf_op} and the normalization condition $\text{Tr}(\tilde C_{11}^{(l,m)} \tilde \varLambda^{(m)}) = \tilde{\t C}_{11}^{(l,m)} \star \Lambda^{(m)} = 1$, we obtain
\begin{equation}
\label{eq:app_Lexpansion}
     4 \text{Tr} \left( \boldsymbol{\alpha}_\mathbb{\Lambda} ^{(l,m)} \tilde C^{(l,m)} \right) = F^{(l)}+ 4 \text{Tr}\left( \boldsymbol{\alpha}^{(m)} \tilde C_{00}^{(l,m)}\right)+ 
 4 \sqrt{F^{(l)}} \text{Re} \left(  \text{Tr} \left( \boldsymbol{\beta}^{(m)} \tilde C_{10}^{(l,m)}\right)\right),
\end{equation}
and consequently
\begin{equation}
\label{eq:app_Lexpansion_bound}
     F^{(l+m)} = F^{(l)} + 4\min_h \max_{\tilde C^{(l,m)}}  \left[ \text{Tr}\left( \boldsymbol{\alpha}^{(m)} \tilde C_{00}^{(l,m)}\right)+ 
  \sqrt{F^{(l)}} \text{Re} \left(  \text{Tr} \left( \boldsymbol{\beta}^{(m)} \tilde C_{10}^{(l,m)}\right)\right) \right]
\end{equation}
The normalization condition was derived using the fact that $\tilde 
{\t C}_{11}^{(l,m)} \in \text{Comb}[(\emptyset, \mathcal{H}_{2l+1} \otimes \mathcal{R}_l),(\mathcal{H}_{2l+2} , \mathcal{H}_{2l+3}),...,(\mathcal{H}_{2(l+m)-2}, \mathcal{H}_{2(l+m)-1})]$ and $\tilde \Lambda^{(m)} \in \text{Comb}[(\mathcal{H}_{2l+1} \otimes \mathcal{R}_l, \mathcal{H}_{2l+2}),(\mathcal{H}_{2l+3}, \mathcal{H}_{2l+4}),...,(\mathcal{H}_{2(l+m)-1}, \emptyset)]$, so all outputs of $\tilde {\t C}_{11}^{(l,m)}$ are inputs of $\tilde \Lambda^{(m)}$ and \textit{vice versa}.

To further simplify the bound, we maximize each term of the RHS of \eqref{eq:app_Lexpansion_bound} independently and use the inequality between the maximum of a sum and the sum of maxima, which leads to \eqref{eq:itmaxmax} from the main text:
\begin{equation}
\label{eq:app_itmaxmax}
F^{(l+m)} \le F^{(l)} + 4 \min_h  \left[  a^{(m)} + \sqrt{F^{(l)}}  b^{(m)} \right],  a^{(m)}  =   \max_{\tilde C^{(m)}} \text{Tr} \left( \boldsymbol{\alpha}^{(m)} \tilde C_{00}^{(m)}   \right),~~b^{(m)} =  \max_{\tilde C^{(m)}}   \text{Re}~ \text{Tr} \left(  \boldsymbol{\beta}^{(m)} \tilde C_{10}^{(m)} \right).
\end{equation}
The bound \eqref{eq:app_itmaxmax} is less tight than \eqref{eq:app_itnew}, but, as we will prove in the next section, there is no difference between bounds generated by \eqref{eq:app_itmaxmax} and \eqref{eq:app_itnew} asymptotically (for large $N$).

For simplicity, starting from \eqref{eq:app_itmaxmax}, we replace $\tilde C^{(l,m)}$ with $\tilde C^{(m)}$ under maximization, since $\tilde C^{(l,m)}$ has the same structure regardless of $l$ . This is possible, when subsequent teeth of a comb $\Lmb_\theta$ are the same, as we assumed from the very beginning in the main text.
Notice that this assumption is not necessary if we just want to derive recursive bounds as in~\eqref{eq:app_itnew}, but it is important to derive an asymptotic bound.
For simplicity and for the sake of deriving an asymptotic bound, let us also adopt convention $l=0$ for naming of spaces $\mathcal{H}$, $\mathcal{R}$, $\mathcal{A}$.
Then we have $\tilde C^{(m)} \in \t{Comb}[(\mathcal{V}_0, \mathcal{R}_0 \otimes \mathcal{H}_1), (\mathcal{H}_2, \mathcal{H}_3),...,(\mathcal{H}_{2m-2} , \mathcal{H}_{2m-1} \otimes \mathcal{A}_m)]$, see Fig.~\ref{fig:app_Lcomb} for a comparison.

\subsection{Derivation of asymptotic bounds \eqref{eq:cor_SS} and \eqref{eq:cor_HS}}


To derive the asymptotic bounds we will use the following lemma.

\vspace{0.5\baselineskip}
\noindent
\textbf{Lemma 1} Let $x^{(n)}$ be a sequence of real numbers satisfying $x^{(n+1)} = x^{(n)} + A + 2 B \sqrt{x^{(n)}}$ for any integer $n$, $x^{(0)}=0$, $A \ge B^2$. Then
\begin{equation}
    \lim_{n \rightarrow \infty } x^{(n)}/n = A ~~\t{for}~~ B=0; \quad \lim_{n \rightarrow \infty } x^{(n)}/n^2 = B^2 ~~\t{for}~~ B \ge 0.
\end{equation}

\vspace{0.5\baselineskip}
\noindent
\textbf{Proof} When $B=0$, then $x^{(n)} = An$, so the first part of lemma is obvious. 
When $B>0$, 
 then for any $n \ge 1$ 
\begin{equation}
\label{eq:app_ai_1}
    x^{(n)} \le f(n) =  An + B^2n(n-1)+ (A-B^2) n \log n, 
\end{equation}
which is proven in the Supplementary Material of Ref.~\cite{Kurdzialek2023a}, Appendix~E, in which $\|\alpha\|$ and $\|\beta\|$ play the role of $A$ and $B$.
Let us now prove that for any $n \ge 1$ 
\begin{equation}
\label{eq:app_ai_3}
    x^{(n)} \ge B^2 n^2.
\end{equation}
This can be shown by induction, we have $x^{(1)} = A \ge B^2 1^2$, and when \eqref{eq:app_ai_3} holds for some $n$, then for $n+1$ we have
\begin{equation}
    x^{(n+1)} \ge B^2 n^2 + 2B^2 n + A \ge B^2 n^2 + 2B^2 n + B^2 =B^2(n+1)^2.
\end{equation}
From \eqref{eq:app_ai_1} , \eqref{eq:app_ai_3}, and the fact that $\lim_{n \rightarrow \infty} f(n)/n^2 = B^2$ follows the second part of the lemma. $\blacksquare$

\vspace{0.5\baselineskip}
The asymptotic bounds for uncorrelated noise \eqref{eq:uncor_SS} and \eqref{eq:uncor_HS} are direct consequences of the iterative bound \eqref{eq:old_it} and Lemma 1---notice that asymptotically (for large $\mathcal{F}_\text{AD}^{(i)}$)  the minimum over $h$ is always achieved when $\|\beta\|$ is minimal, since it is multiplied by $\sqrt{\mathcal{F}_\text{AD}^{(i)}}$ and consequently 
 eventually dominates over the term with $\|\alpha\|$, unless $\|\beta\|=0$.  

 The assumptions of Lemma 1 are true for the uncorrelated case because of the inequality $\| \alpha \| \ge \|\beta\|^2$. Let us show that the analogous inequality holds for the correlated case as well.

 \vspace{0.5\baselineskip}
\noindent
\textbf{Lemma 2} $a^{(m)} \ge {b^{(m)}}^2$

\vspace{0.5\baselineskip}
\noindent
\textbf{Proof} Let us use the notation $\tilde{\t C}^\text{A}$, $\tilde{\t C}^\text{B}$ for combs $ \tilde{\t C}^{(m)}$ maximizing $a^{(m)}$ and $b^{(m)}$ respectively. Because $a^{(m)} = \text{Tr}\left( \boldsymbol{\alpha}^{(m)} \tilde C^\text{A}_{00}\right) \ge \t{Tr}\left( \boldsymbol{\alpha}^{(m)} \tilde C^\t{B}_{00}\right)$ and ${b^{(m)}}^2 \le  \left| \text{Tr} \left( \boldsymbol{\beta}^{(m)} \tilde C_{10}^\text{B} \right)\right|^2$, the inequality
\begin{equation}
    \t{Tr}\left( \boldsymbol{\alpha}^{(m)} \tilde C^\t{B}_{00}\right) \ge \left| \text{Tr} \left( \boldsymbol{\beta}^{(m)} \tilde C_{10}^\text{B} \right)\right|^2
\end{equation}
implies the thesis of our lemma. To prove this inequality, notice that

\begin{multline}
        \left| \text{Tr} \left( \boldsymbol{\beta}^{(m)} \tilde C_{10}^\text{B} \right)\right| = \left| 
 \text{Tr} \left( \text{Tr}_{\text{out}} \left( \sum_k \ket{\dot {\tilde K}_{k,\theta}^{(m)}} \bra{{ K}_{k,\theta}^{(m)}} \right) \tilde C_{10}^\text{B} \right) \right| = \left|\sum_k \braket{{ K}_{k,\theta}^{(m)}|\tilde C_{10}^\text{B} \otimes \mathbb{1}_\text{out}|\dot {\tilde K}_{k,\theta}^{(m)}} \right| = \\ =\left|\sum_k \braket{{ K}_{k,\theta}^{(m)}|  \left( \tilde C_{11}^\text{B}\right)^{\frac{1}{2}} \left(\tilde C_{11}^\text{B} \right)^{-\frac{1}{2}} \tilde C_{10}^\text{B} \otimes \mathbb{1}_\text{out}  |\dot {\tilde K}_{k,\theta}^{(m)}} \right| \le \\ \overset{\text{(i)}}{\le} \sqrt{\sum_k \braket{{ K}_{k,\theta}^{(m)}| \tilde C_{11}^\text{B} \otimes \mathbb{1}_\text{out}| { K}_{k,\theta}^{(m)}}} \cdot \sqrt{\sum_k \braket{\dot {\tilde K}_{k,\theta}^{(m)}|\tilde C_{10}^{\text{B} \dagger}  \left( \tilde C_{11}^\text{B}\right)^{-1} \tilde C_{10}^{\text{B } }\otimes \mathbb{1}_\text{out} | \dot {\tilde K}_{k,\theta}^{(m)}}} \overset{\text{(ii)}}{\le} \sqrt{\sum_k \braket{\dot {\tilde K}_{k,\theta}^{(m)}| \tilde C_{00}^\text{B} \otimes \mathbb{1}_\text{out}|\dot {\tilde K}_{k,\theta}^{(m)}}} = \\ = \sqrt{\text{Tr}\left( \boldsymbol{\alpha}^{(m)} \tilde C^\text{B}_{00}\right)}
\end{multline}
In (i) we used Cauchy-Schwarz inequality $\left| \sum_k \braket{x_k|y_k} \right| \le \sqrt{\sum_k \braket{x_k|x_k}} \sqrt{\sum_k \braket{y_k|y_k}}$ with $\ket{x_k} =  \left( \tilde C_{11}^\text{B}\right)^{\frac{1}{2}} \otimes \mathbb{1}_\text{out} \ket{K_{k,\theta}^{(m)}}$, $\ket{y_k} = \left(\tilde C_{11}^\text{B} \right)^{-\frac{1}{2}} \tilde C_{10}^\text{B} \otimes \mathbb{1}_\text{out}  \ket{\dot {\tilde K}_{k,\theta}^{(m)}}$. In (ii) we used the fact that the expression under the first square root is $\text{Tr} \left(\varLmb_\theta^{(m)} \tilde C_{11}^\text{B}\right) = 1$, since $\varLmb_\theta^{(m)}$ and $\tilde C_{11}^\text{B}$  are two combs whose inputs and outputs are compatible. We also used the fact that the matrix $\tilde C^B = \begin{bmatrix}
           \tilde C^\text{B}_{00} & \tilde C^\text{B}_{01} \\
           \tilde C^\text{B}_{10} &\tilde C^\text{B}_{11}  \\
         \end{bmatrix} $ is hermitian and positive-semidefinite, which implies that $\tilde C^\text{B}_{01} = \tilde C^{\text{B} \dagger}_{10}$, and due to Schur's complement condition $\tilde C_{00}^\text{B} \succeq \tilde C_{10}^{\text{B} \dagger}  \left( \tilde C_{11}^\text{B}\right)^{-1} \tilde C_{10}^{\text{B} }$, which means that $\braket{u|\tilde C_{00}^\text{B} \otimes \mathbb{1}|u} \ge \braket{u|\tilde C_{10}^{\text{B} \dagger}  \left( \tilde C_{11}^\text{B}\right)^{-1} \tilde C_{10}^\text{B} \otimes \mathbb{1}|u} $ for any $\ket{u}$. $\blacksquare$

\vspace{0.5\baselineskip}

If we perform the minimization over $h$ in \eqref{eq:app_itmaxmax} for $F^{(l)} \rightarrow \infty$, then the term with $b^{(m)}$ dominates the term with $a^{(m)}$.
When there is an $h$ for which $b^{(m)} = 0$, then for large enough $l$ the minimum over $h$ in recursive steps is achieved when $b^{(m)} = 0$, and, according to Lemma 1 applied for a sequence $F^{(0)}, F^{(m)}, F^{(2m)},...$, we get
\begin{equation}
\label{eq:app_cor_SS}
    \lim_{N \rightarrow \infty } \frac{\mathcal{F}_\text{AD}^{(N)}}{N} \le \frac{4}{m} \min_h  a^{(m)} ~ \text{s.t.}~~b^{(m)}=0,
\end{equation}
which is \eqref{eq:cor_SS} from the main text. When there is no such $h$, in the large $l$ limit the minimum over $h$ in \eqref{eq:app_itmaxmax} is achieved when $b^{(m)}$ is minimal. Therefore, using  Lemmas 1 and 2  for a sequence $F^{(0)}, F^{(m)}, F^{(2m)},...$ we get
\begin{equation}
    \label{eq:app_cor_HS}
    \lim_{N \rightarrow \infty } \frac{\mathcal{F}_\text{AD}^{(N)}}{N^2} \le \frac{4}{m^2} \min_h  {b^{(m)}}^2.
\end{equation}

Notably, according to Lemma 1, we did not loose any tightness while going from the recursion \eqref{eq:app_itmaxmax} to the asymptotic bounds  \eqref{eq:app_cor_SS} and \eqref{eq:app_cor_HS}.
However, the tightness could be potentially lost when we replaced maximum of sum in \eqref{eq:app_Lexpansion_bound} with sum of maxima in \eqref{eq:app_itmaxmax}.
Let us now demonstrate that this has no effect asymptotically, so after iterating \eqref{eq:app_itnew} many times we get an asymptotic behavior predicted by \eqref{eq:app_cor_SS} or \eqref{eq:app_cor_HS}. 
To show this, let us consider two cases:
\begin{enumerate}
    \item Heisenberg scaling is possible. Then, for $F^{(l)} \rightarrow \infty$ the term $
  \sqrt{F^{(l)}} \text{Re} \left(  \text{Tr} \left( \boldsymbol{\beta}^{(m)} \tilde C_{10}\right)\right)$ dominates over the term $\text{Tr}\left( \boldsymbol{\alpha}^{(m)} \tilde C_{00}\right)$ in the RHS of \eqref{eq:app_Lexpansion_bound} . When we pick $\tilde C^{(l+m)}$ such that $\tilde C _{10}$ maximizes the dominating term, then the result of a  maximization  over $\tilde C^{(l+m)}$ is arbitrarily close to the result of independent maximization over $\tilde C^A$, $\tilde C^B$ for $F^{(l)} \rightarrow \infty$ (the ratio between the two results $\rightarrow 1$).
   \item Heisenberg scaling is not possible.  Then, to perform minimization over $h$ in \eqref{eq:app_Lexpansion_bound}  for  $F^{(l)} \rightarrow \infty$, we should choose $h$ for which $\text{Re} \left(  \text{Tr} \left( \boldsymbol{\beta}^{(m)} \tilde C_{10}\right)\right) = 0$  for any $\tilde C_{10}$. Then, we can choose $\tilde C_{10}=0$ without affecting the result. This makes the condition $\tilde C^{(m)} \succeq 0$ equivalent to $\tilde C_{00}, \tilde C_{11} \succeq 0$ , and the maximization over $\tilde C^{(m)}$ of the sum of terms is equivalent to maximization of the first term over $\tilde C_{00}$.
\end{enumerate}
\subsection{Simpler form of the asymptotic bound for standard scaling}
The asymptotic bound \eqref{eq:app_cor_SS} is expressed as a minimization over all hermitian matrices $h$ for which the constraint $b^{(m)} = 0$ is satisfied.
This constraint is hard to deal with because $b^{(m)}$ is defined as a result of a non-trivial maximization, see \eqref{eq:app_itmaxmax}---let us therefore try to formulate it in a more manageable  way. 

Let $\mathcal{X}_{m}$ be a linear subspace of space $\mathcal{X}_m^{\t{tot}} = \mathcal{L}(\mathcal{R}_0 \otimes \mathcal{H}_1 \otimes \mathcal{H}_2 \otimes ... \otimes \mathcal{H}_{2m-1})$ such that $X \in \mathcal{X}_m$ iff 
\begin{enumerate}
    \item $\text{Tr}(X) = 0$,
    \item There exists a sequence of operators $X^{(1)}, X^{(2)},..., X^{(m-1)}, X^{(m)}$ for which $X = X^{(m)}$, $\forall_{2 \le k \le m} \text{Tr}_{\mathcal{H}_{2k-1}}X^{(k)}=X^{(k-1)} \otimes \mathbb{1}_{\mathcal{H}_{2k-2}}$, $X^{(1)} \in \mathcal{L}(\mathcal{H}_{1} \otimes \mathcal{R}_0)$.

\end{enumerate}

Notice that condition 2. is similar to the linear comb conditions, a space $\mathcal{X}_m$ has a similar structure to the set of CJ operators of combs, but without positivity constraint and with zero trace.  Let us also define $\mathcal{X}_m^\perp$ as an orthogonal complement of $\mathcal{X}_{m}$ in the space $\mathcal{X}_m^{\t{tot}}$ with respect to Hilbert-Schmidt scalar product $\left( A | B\right) = \t{Tr}(A B ^\dagger)$. Since $\boldsymbol{\beta}^{(m)} \in \mathcal{X}_m^\t{tot}$, we can uniquely decompose it as  $\boldsymbol{\beta}^{(m)} = \boldsymbol{\beta}^{(m)}_1 + \boldsymbol{\beta}^{(m)}_2$, where $\boldsymbol{\beta}^{(m)}_1 \in \mathcal{X}_{m}$, $\boldsymbol{\beta}_2^{(m)} \in \mathcal{X}_{m}^\perp$. 

We are now ready to prove the following statement.

\vspace{0.5\baselineskip}
\noindent
\textbf{Lemma 3}  $b^{(m)} \ge 0$, moreover $b^{(m)} = 0$ iff $\bm{\beta}_1^{(m)} = 0$.

\vspace{0.5\baselineskip}
\noindent
\textbf{Proof} 
         The comb conditions for $\tilde C^{(m)}$  are equivalent to the following set of conditions for its blocks $\tilde C^{(m)}_{ij} $: (i) $\tilde C^{(m)}_{00}, \tilde C^{(m)}_{11} \in \text{Comb} [(\emptyset,\mathcal{H}_1 \otimes \mathcal{R}_0 ), (\mathcal{H}_2, \mathcal{H}_3),...,(\mathcal{H}_{2m-2}, \mathcal{H}_{2m-1})]$; (ii) $\tilde C^{(m)}_{01}, \tilde C^{(m)}_{10} \in \mathcal{X}_{m}$; (iii) $\tilde C^{(m)}_{01} = \tilde C^{(m) \dagger}_{10}$ and $\tilde C^{(m)}_{00} \succeq \tilde C^{(m) \dagger }_{10}    \left(\tilde C^{(m)}_{11}\right)^{-1} \tilde C^{(m)}_{10}$. When $\tilde C^{(m)}_{11}$ is singular, then its inverse should be replaced with pseudo inverse. Conditions (i) and (ii) are a consequence 
         of the linear constraints for the comb $\tilde C^{(m)}$, and condition (iii) is a
         consequence of the positivity constraint and Schur's complement condition.
         If we choose any  $\tilde C_{00}^{(m)}$, $\tilde C_{11}^{(m)}$ satisfying (i) and set $\tilde C^{(m)}_{01} = \tilde C^{(m)}_{10} = 0$, then conditions (ii)-(iii) are also satisfied and  $\text{Tr} \left( \boldsymbol{\beta}^{(m)} \tilde C_{10}^{(m)} \right) = 0$ which proves that $b^{(m)} \ge 0$. Since for all cases $\tilde C^{(m) \dagger}_{10} \in \mathcal{X}_{m}$ and $\bm{\beta}_2^{(m)} \in \mathcal{X}_{m}^\perp$, we have $\text{Tr} \left( \tilde C^{(m) }_{10} \bm{\beta}_2^{(m)} \right) = 0$, so $b^{(m)} = 0$ when $\bm{\beta}_1^{(m)} = 0$.
         Let us now assume that  $\bm{\beta}_1^{(m)} \ne 0$, and fix strictly positive definite $\tilde C^{(m)}_{00}$ and $\tilde C^{(m)}_{11}$ satisfying condition (i), and $\tilde C_{01}^{(m)} = \epsilon \bm{\beta}_1^{(m)}$, $\tilde C_{10}^{(m)} = \epsilon \bm{\beta}_1^{(m) \dagger}$.
         Then, condition (ii) is satisfied since $\bm{\beta}_1^{(m)} \in \mathcal{X}_{m}$, and for small enough $\epsilon > 0$ condition (iii) is also satisfied since $\tilde C_{00}^{(m)}$ is strictly positive. Therefore, there exists a comb $\tilde C ^{(m)}$  for which $\text{Re} \left(  \text{Tr} \left( \boldsymbol{\beta}^{(m)} \tilde C_{10}^{(m)} \right) \right) = \epsilon \text{Re} \left(  \text{Tr} \left( \boldsymbol{\beta}_1^{(m)} \boldsymbol{\beta}_1^{(m) \dagger} \right) \right) > 0 $, so $b^{(m)} > 0$, which finishes the proof of the second part of the lemma.    $\blacksquare$

\vspace{0.5\baselineskip}

According to Lemma 3 the condition $b^{(m)} = 0$ can be replaced with condition $\bm{\beta}_1^{(m)} = 0$, or, equivalently, with the condition $\boldsymbol{\beta}^{(m)} \in \mathcal{X}_m^\perp$.

\subsection{Formulations as SDPs}
\label{app:corr_SDP}
To formulate the bounds as SDPs, we will use the following Lemma.

\vspace{0.5\baselineskip}
\noindent
\textbf{Lemma 4} Let us consider the following primal SDP maximization problem:
    \begin{equation}
    \begin{array}{crl}
     \max_{C}  \text{Tr}(A C)  \\
    \mbox{s.t.}& C \in \text{Comb}[(\mathcal{K}_1, \mathcal{K}_2),...(\mathcal{K}_{2N-1}, \mathcal{K}_{2N})] 
\end{array} \ ,
\end{equation}
 $A$ is a hermitian matrix, here $\t{Comb}[...]$ is the set of CJ operators of combs. Then the dual problem is
\begin{equation}
\label{eq:app_dual_lemma}
    \begin{array}{crl}
     \min_{Q^{(1)}, Q^{(2)}, ..., Q^{(N)}}  \text{Tr} \left(Q^{(1)} \right)  \\
    \mbox{s.t.}& Q^{(N)} \otimes \mathbb{1}_{2N }\succeq A \\
    &  \text{Tr}_{2k-1} Q^{(k)} = Q^{(k-1)} \otimes \mathbb{1}_{2k-2}~~\text{for}~~k \in \{2,3,...,N\}
\end{array} \ ,
\end{equation}
where $Q^{(k)}$ are hermitian matrices, $Q^{(k)} \in \mathcal{L} (\mathcal{K}_1 \otimes \mathcal{K}_2 \otimes ... \otimes \mathcal{K}_{2k-1})$.
The optimal value 
of the dual problem is equal to optimal value 
of the primal problem (strong duality).

\vspace{0.5\baselineskip}
\noindent
\textbf{Proof}  A very similar statement was proven in the Supplementary Material of Ref.~\cite{Altherr2021}, see Section I.B, Lemmas 5 and 6.
There,
$A$ was assumed to be equal to performance operator, but this assumption was not used, and the proof remains correct for any hermitian $A$.
Moreover, the proof provided in Ref.~\cite{Altherr2021}
can be only directly applied to cases when $\mathcal{K}_1$ is a trivial one-dimensional space $\emptyset$. To generalize the proof for a general case, it is enough to apply it for a comb with an additional artificial empty teeth, so we consider $C \in \text{Comb}[(\emptyset, \emptyset), (\mathcal{K}_1, \mathcal{K}_2),...,(\mathcal{K}_{2N-1}, \mathcal{K}_{2N})]$ instead of $C \in \text{Comb}[ (\mathcal{K}_1, \mathcal{K}_2),...,(\mathcal{K}_{2N-1}, \mathcal{K}_{2N})]$. Then, we end up with a dual problem in the form \eqref{eq:app_dual_lemma}.$\blacksquare$

\vspace{0.5\baselineskip}
The iterative bound \eqref{eq:app_itnew} can be formulated as an SDP using Algorithm 1 from Ref.~\cite{Altherr2021} directly applied for parameter-dependent comb $\mathbb{\Lambda}_\theta^{(l,m)}$. Let us use the decomposition \eqref{eq:app_Lmb_decomp}, \eqref{eq:app_Lmb_decomp2}, and define $\ket{\dot c_{k,j}^{(l,m)}(h)} = \prescript{~}{ \mathcal{H}_{2(l+m)} \otimes \mathcal{R}_{l+m} }{\braket{j|\dot {\tilde L}_{k,\theta}^{(l,m)}}}$, where $\ket{j}$ is some orthonormal basis of $\mathcal{H}_{2(l+m)} \otimes \mathcal{R}_{l+m}$ (notice that $\ket{\dot c_{k,j}^{(l,m)}(h)}$ depend linearly on the mixing hermitian matrix $h$). The recursive step \eqref{eq:app_itnew}  can be calculated using the following SDP:
\begin{align}
F^{(l+m)} =  4 & \min_{ h, \{Q^{(k)}\}_{k \in \{1,...,m\}}} \text{Tr}\left(Q^{(1)}\right),  \\
& \t{subject to } \\
& A \succeq 0, \\
\nonumber
& \underset{2\le k \le m-1}{\forall} \t{Tr}_{\mathcal{H}_{2(l+k)}}Q^{(k+1)} =  Q^{(k)} \otimes \openone_{\mathcal{H}_{2(l+k)-1}}, \\ &\t{Tr}_{\mathcal{H}_{2l+2}} Q^{(2)}= Q^{(1)} \otimes \openone_{\mathcal{H}_{2l+1} \otimes \mathcal{R}_l},
\end{align}
where
\begin{equation}
A = \left( \begin{array}{c|ccc}
          Q^{(m)} \otimes \openone_{\mathcal{H}_{2(l+m)-1}}& \ket{\dot c^{(l,m)}_{1,1}(h)}  & \hdots &  \ket{\dot c^{(l,m)}_{r,d}(h)} \\ \hline
        \bra{\dot c^{(l,m)}_{1,1}(h)}&  &  &  	 \\
        \vdots&  &  \mathbb{1}_{d r}  &    \\ 
       \bra{\dot c^{(l,m)}_{r,d}(h)} &  &  &            
 \end{array}\right),
\end{equation}
$d$ is the dimension of $\mathcal{H}_{2(l+m)} \otimes \mathcal{R}_{l+m}$, $r$ is the rank of $\mathbb{\Lambda}_\theta^{(l,m)}$, $Q^{(1)} \in \mathcal{L}(\mathcal{V}_l)$, notice that $\ket{\dot c_{k,j}^{(l,m)}(h)}$ depend on $F^{(l)}$.

The asymptotic bound \eqref{eq:app_cor_SS} can be written as a similar SDP with additional constraints for $h$ coming from condition $b^{(m)} = 0$.
First, let us notice that the maximization over $\tilde C^{(m)} $ in $a^{(m)}$ in \eqref{eq:app_cor_SS} boils down to a maximization over $\tilde C_{00}^{(m)} \in \text{Comb}[(\emptyset, \mathcal{H}_{1} \otimes \mathcal{R}_0),(\mathcal{H}_{2}, \mathcal{H}_{3}),...,(\mathcal{H}_{2m-2}, \mathcal{H}_{2m-1})]$. 
Secondly, the condition $b^{(m)} = 0$ can be written as $\boldsymbol{\beta}^{(m)} \in \mathcal{X}_{m}^\perp$ (see Lemma 3 and the remark below).
Since the dual affine space to comb space is another comb space (with outputs and inputs interchanged)  \cite{Bavaresco2021, Liu2023b}, it can be shown that $\boldsymbol{\beta}^{(m)} \in \mathcal{X}_{m}^\perp$ iff there exists a sequence of operators $Y^{ (1)}, Y^{ (2)},...,Y^{ (m-1)}$ for which $\boldsymbol{\beta}^{(m)} =Y^{ (m-1)} \otimes \openone_{\mathcal{H}_{2m-1}}$, $ \forall_{2 \le k \le m-1} \text{Tr}_{\mathcal{H}_{2k}} Y^{(k)} = Y^{(k-1)} \otimes \openone_{\mathcal{H}_{2k-1}}$, $\text{Tr}_{\mathcal{H}_{2}} Y^{(1)} = Y^{(0)} \openone_{\mathcal{H}_{1} \otimes \mathcal{R}_0}$, $Y^{(0)} \in \mathbb{C}$.
By introducing the notation $\ket{ c^{(m)}_{k,j}(h)} = \prescript{~}{\mathcal{H}_{2m} \otimes \mathcal{R}_{m}}{\braket{j| {K}_{k,\theta}^{(m)}}}$,  $\ket{\dot c^{(m)}_{k,j}(h)} = \prescript{~}{\mathcal{H}_{2m} \otimes \mathcal{R}_{m}}{\braket{j|\dot {\tilde K}_{k,\theta}^{(m)}}}$, we get 
\begin{equation}
\label{eq:app_beta}
\boldsymbol{\beta}^{(m)}(h) = \sum_{k,j} \ket{\dot c^{(m)}_{k,j}(h)} \bra{c^{(m)}_{k,j}}, 
\end{equation}
notice that $\boldsymbol{\beta}^{(m)}(h)$ depends linearly on $h$.
Finally, after supplementing the SDP from Ref.~\cite{Altherr2021} with the condition for $\boldsymbol{\beta}^{(m)}$, we get the following SDP for the asymptotic bound:
\begin{align}
\lim_{N \rightarrow \infty} \mathcal{F}^{(N)}_\text{AD} / N \le  4/m & \min_{ h, \{Q^{(k)},Y^{(k)}\}_{k \in \{0,1,...,m-1\}}} Q^{(0)},  \\
& \t{subject to } \\
& A \succeq 0, \\
\nonumber
& \underset{2\le k \le m-1}{\forall} \t{Tr}_{\mathcal{H}_{2k}}Q^{(k)} =  Q^{(k-1)} \otimes \openone_{\mathcal{H}_{2k-1}}, \\ &\t{Tr}_{\mathcal{H}_{2}} Q^{(1)}= Q^{(0)} \openone_{\mathcal{H}_1 \otimes \mathcal{R}_{0}},~~Q^{(0)} \in \mathbb{R}, \\
&\boldsymbol{\beta}^{(m)}(h) =Y^{ (m-1)} \otimes \openone_{\mathcal{H}_{2m-1}}, \\
&\forall_{2 \le k \le m-1} \text{Tr}_{\mathcal{H}_{2k}} Y^{(k)} = Y^{(k-1)} \otimes \openone_{\mathcal{H}_{2k-1}}, \\
&\text{Tr}_{\mathcal{H}_{2}} Y^{(1)} = Y^{(0)} \openone_{\mathcal{H}_{1} \otimes \mathcal{R}_0},~~Y^{(0)} \in \mathbb{C},
\end{align}
where 
\begin{equation}
A = \left( \begin{array}{c|ccc}
          Q^{(m-1)} \otimes \openone_{\mathcal{H}_{2m-1}}& \ket{\dot c^{(m)}_{1,1}(h)}  & \hdots &  \ket{\dot c^{(m)}_{r,d}(h)} \\ \hline
        \bra{\dot c^{(m)}_{1,1}(h)}&  &  &  	 \\
        \vdots&  &  \mathbb{1}_{d r}  &    \\ 
       \bra{\dot c^{(m)}_{r,d}(h)} &  &  &            
 \end{array}\right),
\end{equation}
$d$ is dimension of $\mathcal{H}_{2m} \otimes\mathcal{R}_{m} $, $r$ is the rank of $\varLmb_\theta^{(m)}$.

The asymptotic bound in the presence of HS \eqref{eq:app_cor_HS} can be written as
\begin{equation}
     \lim_{N \rightarrow \infty } \mathcal{F}_\text{AD}^{(N)}/N^2 \le \left[ 2/m \min_h \max_{\tilde C^{(m)} \in \text{Comb}[(\mathcal{V}_0, \mathcal{H}_1 \otimes \mathcal{R}_0 ),(\mathcal{H}_2, \mathcal{H}_3),...,(\mathcal{H}_{2m-2}, \mathcal{H}_{2m-1}) ]}  \text{Tr} \left(\tilde C^{(m)} \begin{bmatrix}
           0 & \frac{1}{2} \boldsymbol{\beta}^{(m)} \\
           \frac{1}{2} \boldsymbol{\beta}^{(m)\dagger} &0  \\
         \end{bmatrix} \right) \right]^2,
\end{equation}
where we used the block decomposition $\tilde C^{(m)} =\begin{bmatrix}
           \tilde C_{00}^{(m)} & \tilde C_{01}^{(m)} \\
           \tilde C_{10}^{(m)} & \tilde C_{11}^{(m)}  \\
         \end{bmatrix} $ where $\tilde C_{01}^{(m)\dagger} = \tilde C_{10}^{(m)}$. After dualizing the maximization problem using Lemma 4, we can write the whole min max problem as a single SDP minimization:
\begin{align}
\sqrt{\lim_{N \rightarrow \infty } \mathcal{F}_\text{AD}^{(N)}/N^2} \le 2/m & \min_{h,\{Q^{(k)}\}_{k \in \{1,...,m\}}} \text{Tr} \left( Q^{(1)} \right),\\ & \t{subject to } \\ 
&Q^{(m)} \otimes \openone_{\mathcal{H}_{2m-1}} \succeq \begin{bmatrix}
           0 & \frac{1}{2} \boldsymbol{\beta}^{(m)} \\
           \frac{1}{2} \boldsymbol{\beta}^{(m)\dagger} &0  \\
         \end{bmatrix}, \\
\nonumber
& \underset{2\le k \le m-1}{\forall} \t{Tr}_{\mathcal{H}_{2k}}Q^{(k+1)} =  Q^{(k)} \otimes \openone_{\mathcal{H}_{2k-1}}, \\ &\t{Tr}_{\mathcal{H}_{2}} Q^{(2)}= Q^{(1)} \otimes \openone_{\mathcal{R}_0 \otimes \mathcal{H}_{1}},
\end{align}
where $Q^{(1)} \in \mathcal{L}(\mathcal{V}_0)$, $\boldsymbol{\beta}^{(m)}$ is given by \eqref{eq:app_beta}.
\section{Examples}
The code we used to generated upper bounds introduced in this work and tensor-network based lower bounds introduced in \cite{Kurdzialek2024, Chabuda2020} is available in a public repository \cite{dulian_package}.
To generate upper bounds, we used the function \texttt{ad\_asym\_bound\_correlated} from file \texttt{qmetro/bounds.py}.
To generate lower bounds for correlated dephasing examples, we used tensor-network based optimization over adaptive strategies with limited ancilla dimension $d_\mathcal{A}$, this was done using the function \texttt{iss\_tnet\_adaptive\_qfi} from the file \texttt{qmetro/protocols/iss.py}.
We calculated lower bounds for the collisional thermometry example by considering a parallel scheme, when input state and output measurement are represented using tensor networks---the optimization was performed using the function \texttt{iss\_tnet\_parallel\_qfi} from the file \texttt{qmetro/protocols/iss.py}.

\subsection{Correlated dephasing}
\label{appE1}
All the details related to the correlated dephasing model are described in Ref.~\cite{Kurdzialek2024} (note that here we use notation $S$ instead of $T$ for a mixing map).
The only additional difficulty is that the tightness of an upper bound depends on the way in which we cut the chain of correlated channels into pieces.
As we checked numerically, for parallel dephasing the tightest bound is obtained when we decompose mixing map as $S = \sqrt{S} \circ \sqrt{S}$, and make cuts between two maps $\sqrt{S}$, see Fig. 2(a) in the main text.
The Kraus operators of channel $\t S$ implementing a stochastic map $S_{i|i-1}$ on the basis $\ket{\pm}$ are  $S_{sr} =\sqrt{\frac{1+srC}{2}} \ket{s} \bra{r} $ for $s, r \in \{+,-\} $. It can be shown by direct calculations, that the square root map for $C>0$ is
\begin{equation}
    \sqrt{S}_{i|i-1}(r_i | r_{i-1}) = (1 + r_i r_{i-1} \sqrt{|C|})/2,
\end{equation}
the corresponding Kraus operators of a quantum map are $\sqrt{S}_{sr} =\sqrt{\frac{1+sr \sqrt{|C|}}{2}} \ket{s} \bra{r} $for $s, r \in \{+,-\} $. When $C<0$, we should construct a map like for positive $C$ of the same absolute value, but then additionally add a flip between elements of basis $\ket{+}$, $\ket{-}$ between two maps $\sqrt{S}$.
\subsection{Collisional thermometry}
\label{appE2}
The collisional quantum thermometry model was introduced in Ref.~\cite{Seah2019}. The thermal bath of temperature $T$ is probed by a qubit thermometer whose hamiltonian is $H = \hbar \Omega \sigma_z /2 $.
The thermalization of a thermometer is described by a quantum master equation in Lindblad form
\begin{equation}
    \frac{\t{d} \rho}{\t{d}t} = \mathcal{L}_T(\rho) =  \gamma (\bar n +1) \mathcal{D}(\sigma_-) + \gamma \bar n \mathcal{D}(\sigma_+),
\end{equation}
where $\mathcal{D}(X) = X \rho X^\dagger - \frac{1}{2} \{X^\dagger X, \rho\} $, $\sigma_\pm = (\sigma_x \pm i \sigma_y)$, $\bar n = (e^{\hbar \Omega / k_B T} -1)^{-1} $, $\gamma$ is a coupling constant between the thermometer and the thermal bath.
The channel describing the thermalization for a time $\tau$ is given by integrating the above equation
\begin{equation}
    \t S^\tau_T (\rho)= e^{\mathcal{L}_T \tau} (\rho),
\end{equation}
which is a generalized amplitude damping channel.
When $\tau \rightarrow \infty$, then the output of a channel $\t S_T^\tau$ is a Gibbs state $\rho_{\t{th}, T} = e^{-H_\mathcal{E} / k_B T} / Z$, the QFI of this state is a thermal Fisher information $F_{\t{th},T} = F( \rho_{\t{th}, T})$.
Interestingly, the channel QFI of $\t S_T^\tau$ can be larger than $F_{\t{th},T}$. We numerically study a case when $k_B T / \hbar \Omega = 2$, which means that $\bar n = 1.514$ (the same value of $\bar n$ was chosen in Ref.~\cite{Seah2019}). We observed that the maximal channel QFI is achieved for $\tau_\t{opt} \approx 0.417 \gamma^{-1}$, then $\mathcal{F}(\t S _T^\tau) \approx 5.66 F_{\t{th},T} $, see Fig.~\ref{fig:app_coll_plots}(a).
Notably, one needs to entangle the input state with ancilla to achieve this optimal channel QFI.

\begin{figure}[h]
\includegraphics[width=0.45\columnwidth]{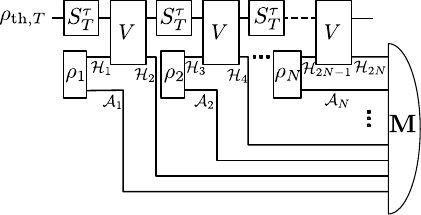}
\caption{Collisional quantum thermometry: thermalizing channels $\t S _T ^\tau$ acting on thermometer are interwined with unitary channels $V$ describing energy exchange between mediators ($\mathcal{H}$) and thermometer. At the end, the joint measurement $\mathbf{M}$ is performed on all mediators and ancillary qubits (initially entangled with mediators). Our upper bounds cover strategies with all possible entangled states of mediators and ancillas. We also calculate lower bounds for the strategy in which initially mediators are not entangled with each other, but each mediator ($\mathcal{H}_{2i-1}$) can be entangled with the corresponding ancilla ($\mathcal{A}_i$).}
\label{fig:app_coll_scheme}
\end{figure}

In a collisional scheme, there is no direct access to a thermometer, one only has access to mediator qubits that exchange energy with thermometer after each thermalization process, the interaction between thermometer and mediator is governed by the unitary $V = e^{-i g t (\sigma_+ \otimes \sigma_- + \sigma_- \otimes \sigma_+)}$, $g$ is a coupling constant, $t$ is an interaction time; a full energy swap occurs for $gt = 0.5 \pi$, in this work we numerically analyze the case $gt = 0.35 \pi$. In Fig.~\ref{fig:app_coll_scheme}, we present a collisional thermometry scheme, in which all mediators are prepared in a product state, but each mediator is assisted with a qubit ancilla, which can be entangled with it. Mediators interact with thermometer, and at the end, all mediators and ancillary qubits are measured. Note, that the thermometer is initialized in a thermal state, whereas in Ref.~\cite{Seah2019} the initial state of a thermometer is a stationary state of concatenation of maps $V$ and $\t S_T^\tau$. This stationary state depends on the input mediator of $V$, so we cannot fix a stationary state when computing universal bounds. However, the choice of initial state of thermometer does not affect the asymptotic results.

Importantly, our upper bounds cover a much broader class of strategies, in which mediators can be prepared in an arbitrary entangled state, and adaptive control between subsequent probings can be applied.

To complement
these general upper bounds, we calculate lower bounds using tensor network techniques~\cite{Chabuda2020, dulian_package}, assuming that the input state of mediators is a product state, each mediator is entangled with an ancillary qubit, and that the final state of mediators and ancillas is measured using a collective measurement $\mathbf{M}$, which is described by a matrix product operator with bound dimension $4$ (which means that some non-locality in a final measurement is allowed).
As demonstrated in Fig. 2(c) in the main text, this strategy is optimal (up to numerical precision) for $\tau = 10 \tau_\t{opt}$, when correlations between subsequent channels are negligible.
For $\tau = \tau_\textrm{opt}$, there is some potential advantage from using entangled states---however, this potential advantage is relatively small.

To make the bounds as tight as possible, we need to judiciously cut the whole chain of channels $V$ and $\t S_T^\tau$ ($\Lambda_\theta^{(N)}$) into smaller pieces $\Lmb_\theta^{(m)}$. More precisely, we need to find combs $\Lmb_\theta^{(m)}$ such that their link product is $\Lambda_\theta^{(N)}$  (environmental spaces of neighbouring combs are linked).
Therefore, we can concatenate $\Lmb_\theta^{(m)}$ with some map $\t X$ acting on its output environment and the inverse of this map $\t X^{-1}$ acting on an input environment---then, after linking $\Lmb_\theta^{(m)}$ with each other, each $\t X$ will be concatenated with $\t X^{-1}$, so we will obtain the same resulting comb $\Lambda_\theta^{(N)}$.
However, the bound tightness might be affected by the choice of $\t X$.
Importantly, $\t X$ must chosen such that the resulting $\Lmb_\theta^{(m)}$ are still valid combs (in particular, they must be completely positive).

In our example we choose $\t X = \t S_{T_0}^{\tau_0}$, which is a map that does not differ from $\t S_{T}^{\tau_0}$ at the value of temperature around which the estimation is performed, but $\frac{\t d}{\t d T} \t S_{T_0}^{\tau_0} = 0 $, which means that $\t S_{T_0}^{\tau_0}$ is insensitive to the changes of $T$, so its channel QFI is $0$. The intuition behind this choice is that such $T$-insensitive thermalization may hide some information that leaks from environment when we cut the whole chain of channels into pieces. 

Obviously, the inverse map $\t S_{T_0}^{- \tau_0}$ is not completely positive. However, the total map acting on an input environment is $ \t S_{T}^{ \tau}\circ \t S_{T_0}^{- \tau_0}$ is completely positive assuming that $\tau_0 < \tau$---the resulting channel at $T = T_0$ is a thermalization with time $\tau - \tau_0$, the channel is full-rank, so it will remain completely positive in the first order expansion irrespectively of its derivative. 

The large values of $\tau_0$ allow to reduce the role of environmental leakages at the outpus of $\Lmb_\theta^{(m)}$, but at the same time, extra anti-thermalization map $\t S_{T_0}^{- \tau_0}$ at the beginning may increase the QFI. Therefore, one need to optimize over $\tau_0$ (assuming $\tau_0 < \tau$) to get the tightest possible bound. Interestingly, for $\tau = \tau_\t{opt}$ the optimal value is $\tau_0 = 0$, so it is not advantagous to use the described trick. The situation is completely different for $\tau = 10 \tau_\t{opt}$, see Fig.~\ref{fig:app_coll_plots}(b), then choosing $\tau_0 > 0$ can tighten the bound significantly. With the optimal value $\tau_0 \approx 0.62 \gamma^{-1} $, we get an upper bound that coincides with lower bounds already for $m=1$.
\begin{figure}[h]
\includegraphics[width=0.8\columnwidth]{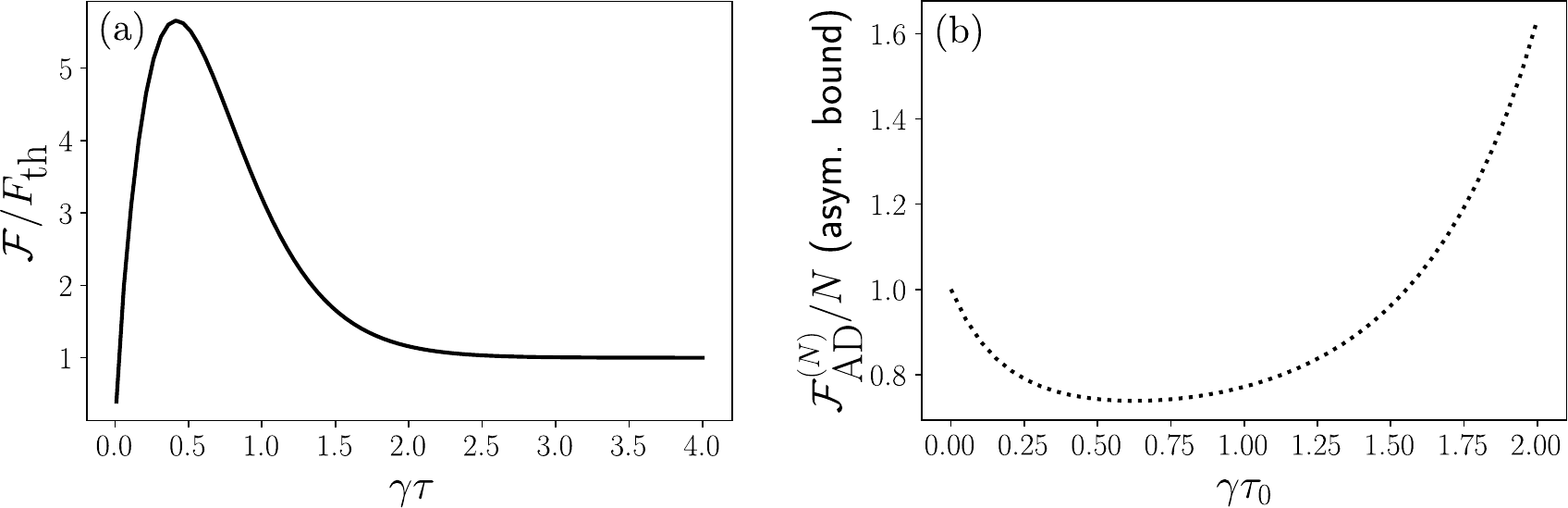}
\caption{ (a): The ratio between channel QFI of a thermal channel $\t S^\tau_T$ and the QFI of a thermal state as a function of thermalization time multiplied by a coupling constant $\gamma \tau$. (b) The tightness of the bound is affected by the choice of time $\tau_0$ in an inverse map $\t S_T^{-\tau_0}$. Here we show how an asymptotic bound depends on $\gamma \tau_0$ for $\tau = 10 \tau_\textrm{opt}$ $m=1$. }
\label{fig:app_coll_plots}
\end{figure}

\end{document}